\documentclass[iop,twocolappendix]{emulateapj}
\usepackage{todonotes}
\usepackage{graphicx}
\usepackage[caption=false]{subfig}
\usepackage{color}
\usepackage{hyperref}
\usepackage{csvsimple}
\usepackage{forloop}

\setcitestyle{aysep={}, citesep={;}}
\definecolor{darkblue}{rgb}{0.,0.,0.8}
\definecolor{darkred}{rgb}{0.5,0.,0.}
\hypersetup{colorlinks=true, linkcolor=darkblue, citecolor=darkred, urlcolor=darkblue}

\newcommand{\soft}[1]{{\scshape{#1}}}

\newcommand{\eg}{e.g.,}
\newcommand{\ie}{i.e.,}
\newcommand{\magarc}{mag arcsec$^{-2}$}

\newcommand{\eqn}[1]{Equation~\ref{eqn:#1}}
\newcommand{\fig}[1]{Figure~\ref{fig:#1}}
\newcommand{\sect}[1]{\S\ref{sec:#1}}
\newcommand{\tab}[1]{Table~\ref{tab:#1}}

\newcommand{\zsim}[1]{\ensuremath{z \simeq {#1}}}
\newcommand{\zeq}[1]{\ensuremath{z = {#1}}}
\newcommand{\LSun}{\ensuremath{L_{\odot}}}
\newcommand{\MSun}{\ensuremath{M_{\odot}}}

\thickmuskip=\medmuskip
\medmuskip=\thinmuskip

\makeatletter
\@ifclassloaded{emulateapj}{\newcommand{\flexstar}{*}}{\newcommand{\flexstar}{}}
\@ifclassloaded{emulateapj}{\newcommand{\rotate}{}}{}
\@ifclassloaded{emulateapj}{}{\newcommand{\LongTables}{}}
\makeatother

\newcommand{\cutoffrank}{30}
\newcommand{\fracqso}{0.39}
\newcommand{\fracgal}{0.30}
\newcommand{\signifprob}{0.78}
\newcommand{\signifsigma}{0.78}

\newcommand{\qsomergers}{7}
\newcommand{\qsononmergers}{11}
\newcommand{\qsonondetects}{1}

\newcommand{\galmergers}{24}
\newcommand{\galnonmergers}{56}
\newcommand{\galnondetects}{4}

\newcommand{\change}[1]{{#1}}

\begin{document}
	
\title{Do the Most Massive Black Holes at \zeq{2} Grow via Major Mergers?\footnote{Based on observations made with the NASA/ESA Hubble Space Telescope, which is operated by the Association of Universities for Research in Astronomy, Inc., under NASA contract NAS 5-26555. These observations are associated with Program ID 12613.}}
\shorttitle{Are Massive Quasars at \zeq{2} Merger-Driven?}

\author{
M. Mechtley\altaffilmark{1,2}, 
K. Jahnke\altaffilmark{1},
R. A. Windhorst\altaffilmark{2},
R. Andrae\altaffilmark{1},
M. Cisternas\altaffilmark{3},
S. H. Cohen\altaffilmark{2},
T. Hewlett\altaffilmark{4},
A. M. Koekemoer\altaffilmark{5}, 
M. Schramm\altaffilmark{6},
A. Schulze\altaffilmark{6},
J. D. Silverman\altaffilmark{6},
C. Villforth\altaffilmark{4},
A. van der Wel\altaffilmark{1}, 
L. Wisotzki\altaffilmark{7}
}

\altaffiltext{1}{Max-Planck-Institut f{\"u}r Astronomie, K{\"o}nigstuhl 17, D-69117, Heidelberg, Germany}
\altaffiltext{2}{School of Earth and Space Exploration, Arizona State University, P.O. Box 871404, Tempe, AZ 85287-1404, USA}
\altaffiltext{3}{Instituto de Astrof\'{\i}sica de Canarias, E-38205 La Laguna, Tenerife, Spain}
\altaffiltext{4}{Scottish Universities Physics Alliance (SUPA), University of St Andrews, School of Physics and Astronomy, North Haugh, St Andrews, KY16 9SS, UK}
\altaffiltext{5}{Space Telescope Science Institute, 3700 San Martin Dr, Baltimore, MD 21218, USA}
\altaffiltext{6}{Kavli Institute for the Physics and Mathematics of the Universe, Todai Institutes for Advanced Study, the University of Tokyo, Kashiwa, Japan 277-8583 (Kavli IPMU, WPI)}
\altaffiltext{7}{Leibniz-Institut f\"ur Astrophysik Potsdam (AIP), An der Sternwarte 16, 14482 Potsdam, Germany}

\shortauthors{Mechtley et al.}

\begin{abstract}
The most frequently proposed model for the origin of quasars holds that the high accretion rates seen in luminous active galactic nuclei are primarily triggered during major mergers between gas-rich galaxies. While plausible for decades, this model has only begun to be tested with statistical rigor in the past few years. Here we report on a Hubble Space Telescope study to test this hypothesis for \zeq{2} quasars with high super-massive black hole masses ($M_\mathrm{BH}=10^9-10^{10}~\MSun{}$), which dominate cosmic black hole growth at this redshift. We compare Wide Field Camera 3 $F160W$ (rest-frame $V$-band) imaging of 19 point source-subtracted quasar hosts to a matched sample of 84 inactive galaxies, testing whether the quasar hosts have \change{greater evidence for} strong gravitational interactions. \change{Using an expert ranking procedure, we find that the quasar hosts are uniformly distributed within the merger sequence of inactive galaxies, with no preference for quasars in high-distortion hosts. Using a merger/non-merger cutoff approach}, we recover distortion fractions of $f_\mathrm{m,qso}=\fracqso{}\pm{}0.11$ for quasar hosts and $f_\mathrm{m,gal}=\fracgal{}\pm{}0.05$ for inactive galaxies (distribution modes, 68\% confidence intervals), with both measurements subjected to the same observational conditions and limitations. The slight enhancement in \change{distorted fraction} for quasar hosts over inactive galaxies is not significant, with a probability that the quasar fraction is higher of $P(f_\mathrm{m,qso}>f_\mathrm{m,gal}) = \signifprob{}$ ($\signifsigma{}\,\sigma$), in line with results for lower mass and lower $z$ AGN. We find no evidence that major mergers are the primary triggering mechanism for the massive \change{quasars} that dominate accretion at the peak of cosmic quasar activity.
\end{abstract}
\maketitle

\section{Introduction}
\label{sec:intro}
How do active galactic nuclei (AGN) get the gas that fuels black hole growth? Proximal to the black hole, the direct mechanism for feeding is described by the unification model for AGN \citep[\eg{}][]{antonucci_unified_1993}: gas forms a thin accretion disk as it falls into a super-massive black hole (SMBH), with gravitational energy converted into kinetic energy that heats the gas, producing the UV/optical continuum that ultimately drives other observed properties, such as broad and narrow emission lines, and re-emission in the infrared via dust heating. The observational evidence for this general model is secure, with most work now focusing on its details. Less well-understood are the physical processes that result in the transport of gas from kiloparsec-scale reservoirs in the galaxy into the central few parsecs, such that it can be captured by the SMBH and accreted. The inferred active-phase lifetimes of $\simeq{} 10^7-10^8$~yr \citep[\eg{}][]{martini_quasar_2001, yu_observational_2002, shen_clustering_2007} mean that galaxies must provide a significant fraction of the final SMBH mass (roughly 1/$e$ if accreting near the Eddington rate) of gas to the very small volume surrounding the SMBH over a comparable timescale. This requires gas transport mechanisms that efficiently strip angular momentum, allowing gas to pass close enough to the SMBH for capture and accretion.

\change{As the most luminous, massive AGN, the gas transport mechanism most often posited for quasars} is disruption due to gravitational interactions with massive galaxies, in particular via major mergers. The most popular version of this model, originally described by \citet{sanders_ultraluminous_1988}, holds that quasars are a phase in galaxy evolution that follows a starburst phase triggered during a gas-rich major merger. This is an entirely plausible scenario, given that the necessary ingredients are present: galaxy-scale torques and large gas reservoirs. \change{Major mergers --- with or without a starburst phase --- almost certainly account for some fraction of quasar triggering.} However, observational evidence that supports this feeding mechanism uniquely \change{as dominant} over other mechanisms (\eg{} violent disk instabilities or direct accretion of intergalactic medium via cold gas streams) has remained elusive.

With high spatial resolution, a relatively stable point spread function, and sensitivity to low surface brightness features in distant galaxies, the Hubble Space Telescope (HST) has been the observatory of choice for many quasar host studies, especially those examining host galaxy morphology. Some early studies of quasar host galaxies with HST lent credence to the major merger model, noting the presence of merging signatures and close companions in some quasar hosts. However, these studies had explicitly biased or unknown sample selection functions, which prevent the results from being easily generalized to the parent populations. For example, \citet{disney_interacting_1995} and \citet{bahcall_hubble_1997} both contain a mix of radio-loud and radio-quiet objects, but with much higher fractions of radio-loud objects than the quasar population generally. While such studies were important in unambiguously resolving host galaxy structures for the first time and helping to understand what kinds of galaxies \emph{can} host quasars, they did not provide conclusive evidence that mergers are the unique or even a dominant trigger for quasar activity.

\change{Later studies began to select more representative quasar samples, with better control over factors like luminosity and redshift ranges. At low redshift --- $z\lesssim{}0.4$, well after the peak of cosmic quasar activity --- the host galaxies of the most luminous quasars were found to be mostly giant ellipticals undergoing only minor (if any) interactions \citep{mclure_comparative_1999, dunlop_quasars_2003, floyd_host_2004}. These studies also took the crucial step of comparing the quasar hosts to inactive galaxies with similar properties \citep{dunlop_quasars_2003}, noting that such small-scale disturbances are a common feature of giant elliptical galaxies, including brightest cluster galaxies. These also included the first HST studies of rest-frame visible light in high-redshift quasar hosts \citep[\eg{} \zeq{1-3},][]{kukula_nicmos_2001, ridgway_nicmos_2001}, though these were generally limited to lower-luminosity quasars and did not systematically examine morphologies.}

\change{Practical considerations have limited most quasar host studies with HST to a few tens of objects each, so implicit or explicit selection biases play an important role in determining how generally a study's conclusions may be applied. We discuss possible biases in our own sample in \sect{biases} below. Many successful studies have focused on specific areas of quasar parameter space, allowing inferences to be made about specific classes of objects. For instance, almost universal evidence for merging hosts has been found in studies using various ``red'' or (semi-)obscured quasar selection methods \citep[\eg{}][]{canalizo_quasistellar_2001, urrutia_evidence_2008, glikman_major_2015}. There is some question whether these represent an evolutionary stage or a subset of quasars with dusty, ULIRG-like hosts, which we discuss in greater detail in \sect{comp_red} below. \citet{chiaberge_radio_2015} also recently examined the different merger fractions of various radio-selected AGN samples, finding significantly higher merger fractions for both faint and luminous radio-loud AGN than for radio-quiet AGN or inactive galaxies. In this study we choose not to select explicitly based on radio-loudness, which results in an implicit focus mainly on radio-quiet AGN, as these make up $\simeq$90\% of the luminous quasar population \citep[\eg{}][]{jiang_radioloud_2007}.}

Comparison to a sample of inactive galaxies is also key to demonstrating that an observed merger fraction is actually related to AGN activity. Large samples of inactive galaxies are observed at all merger stages, so it is clear that a major merger alone is not a sufficient condition for quasar activity. Thus, to conclusively demonstrate that mergers are an important channel for \change{quasar} fueling, we would need to observe an \emph{enhancement} to the merger fraction in \change{quasar} hosts \emph{relative} to a matched sample of inactive galaxies. Several studies of \change{lower-luminosity} AGN host morphologies with inactive control samples have been conducted in HST extragalactic survey fields \citep[\eg{}][]{grogin_agn_2005, gabor_active_2009, cisternas_bulk_2011, schawinski_hst_2011, kocevski_candels:_2012, bohm_agn_2013, villforth_morphologies_2014}. In particular, we designed our study methodology following \citet{cisternas_bulk_2011}, who used visual classification to compare strong distortion signatures in \change{moderate-luminosity} X-ray selected AGN hosts to a comparison sample of inactive galaxies in the redshift range \zeq{0.3-1.0}. They found no significant enhancement to the merger fraction of AGN hosts relative to inactive galaxies, demonstrating that the majority of cosmic black hole mass accretion at $z<1.0$, \ie{} in AGN with inferred SMBH masses $\simeq{} 10^8$~\MSun{} \citep{vestergaard_mass_2009}, is \emph{not} merger-driven. How do we then reconcile this result with the results from the red quasar and radio galaxy studies? One possibility is that certain sub-classes of AGN may be preferentially merger-driven, even though the bulk of all objects are not. In particular, a downsizing trend has been observed, such that near the peak of quasar activity at \zeq{2}, higher-mass SMBHs dominate the cosmic mass accretion \citep[$\simeq{} 10^{9.5}$~\MSun{},][]{vestergaard_mass_2009}. It is possible that forming these most massive black holes requires major mergers, as a particularly efficient gas transport mechanism, and that the declining major merger rate of galaxies is one of the driving forces behind this downsizing trend.

In this paper, we \change{examine the evidence for major mergers} among the hosts of 19 of the highest-mass Type-1 quasars at \zeq{2} --- \ie{} broad-line quasars with $M_\mathrm{BH}=10^{9}-10^{10}$~\MSun{} \change{that dominate cosmic accretion at this redshift \citep{vestergaard_mass_2009}} --- compared to a matched sample of 84 inactive galaxies at the same redshift. This is near the peak of cosmic quasar activity and is the highest redshift where the rest-frame optical emission that reliably diagnoses recent merger signatures can be observed with HST. In \sect{obs}, we describe the selection of the 19 quasars and our observations with the Wide Field Camera 3 (WFC3) infrared channel. In \sect{psf_sub}, we describe the method for modeling and subtracting the central point source from the quasar images. In \sect{comparison_sample}, we describe the selection of the comparison sample of inactive galaxies. In \sect{ranking_and_merger_fraction} we describe the procedure for producing a list of galaxies ranked by \change{evidence for strong gravitational distortions}, based on visual inspection by 10 professionals. Finally, in \sect{discussion}, we discuss our statistical analysis of the ranked list and the resultant merger fractions, as well as properties of the quasar hosts such as inferred galaxy masses. Throughout the paper, we adopt a $\Lambda{}$CDM cosmology with $H_0 = 67$~km~s$^{-1}$~Mpc$^{-1}$, $\Omega{}_M = 0.3$, and $\Omega{}_\Lambda{} = 0.7$ \citep{planck_collaboration_planck_2014}. Unless otherwise stated, all magnitudes are on the AB system \citep{oke_secondary_1983} and have been corrected for Galactic extinction using the reddening map of \citet{schlafly_measuring_2011}.

\section{Observations and Data Reduction}
\label{sec:obs}
\subsection{Sample Definition and Existing Data}
\label{sec:sdss_data}
As discussed in \sect{intro}, a well-defined selection function is necessary to understand the statistical biases present in any AGN sample. Previous contiguous-field morphological studies with HST \citep[\eg{}][]{grogin_agn_2005, gabor_active_2009, cisternas_bulk_2011, kocevski_candels:_2012} are substantially volume-limited and thus lacking luminous \change{quasars} at the highest black hole masses ($M_\mathrm{BH}>10^9$~\MSun{}). We are thus interested in extending these studies to test whether black hole accretion is merger-driven in the highest-mass black holes at the highest redshift where HST studies can reliably diagnose merger signatures --- \zsim{2}, where the rest frame optical emission shifts into the WFC3 IR $F160W$ filter.

We selected quasars from the SDSS 5th Data Release Quasar Catalog \citep{schneider_sloan_2007}, using virial black hole masses calculated from the \ion{Mg}{2} line by \citet{shen_biases_2008}. We required that the quasars be targeted using the uniform color selection of SDSS \citep[\texttt{TARGET\_QSO\_CAP} or \texttt{TARGET\_QSO\_SKIRT}, see][]{richards_spectroscopic_2002, richards_sloan_2006}, have a redshift in the range \zeq{1.9-2.1}, and SMBH mass $M_\mathrm{BH}=10^{9.3}-10^{9.7}$~\MSun{}. In order to ensure an unbiased sample of all optically-luminous massive AGN, the quasars were selected blindly from the parent sample, with no further criteria based on spectral features, broad-band colors, or detections at other wavelengths. Of this parent sample, we submitted 115 randomly-drawn targets for an HST Cycle 19 SNAPshot survey (Program SNAP 12613, PI: Jahnke). Between October 2011 and September 2012, 19 of these quasars were observed with the WFC3 infrared channel in the $F160W$ filter (rest-frame $V$-band). After the HST observations were completed, we examined existing data from the SDSS spectra and wide-area surveys. These data are summarized in \tab{sdss_data} and discussed below.

\citet{shen_catalog_2011} classified the spectral properties of all the SDSS Data Release 7 quasars \citep{schneider_sloan_2010}, including the 19 in our HST program. In \tab{sdss_data} we reproduce those data relevant to deriving the \ion{Mg}{2} virial BH masses --- rest-frame luminosity at 3000\AA{} and FWHM of the broad \ion{Mg}{2} component. We note that \citet{shen_catalog_2011} also publish a revised calibration for \ion{Mg}{2} virial BH masses, different from the \citet{mclure_cosmological_2004} calibration used by \citet{shen_biases_2008} that went into our mass selection. The new calibration simply increases the scatter in BH masses --- the sample is still representative of luminous AGN with the high BH masses and the median mass, $M_\mathrm{BH}=10^{9.5}$~\MSun{}, remains unchanged. \tab{sdss_data} also includes an estimate of the Eddington ratio for each quasar, calculated from the \citet{shen_catalog_2011} bolometric luminosity and \ion{Mg}{2} BH mass\footnote{The Eddington ratio estimate in the \citet{shen_catalog_2011} catalog (\texttt{LOGEDD\_RATIO}) is calculated from the \ion{C}{4} black hole masses for quasars with $z\geq{}1.9$. The Eddington ratio we report is calculated using the \ion{Mg}{2} black hole masses.}. One object was flagged as a \ion{C}{4} broad absorption line (BAL) quasar. We manually examined the SDSS spectra of all 19 objects and confirmed that it is the only BAL quasar in the sample. Eighteen of the quasars were covered by the FIRST 1.4~GHz radio survey \citep[version 14Dec17]{white_catalog_1997}, with a detection limit sufficient to assess radio-loudness --- one object is radio-loud \citep[$S_{1.4GHz}=900$~mJy, $R=1670$, core-dominated, see classification scheme of][]{jiang_radioloud_2007}. Additionally, one object is classified as hot dust-poor, defined as having a rest-frame mid-IR flux deficiency \citep{jiang_dust-free_2010, jun_physical_2013}, with a $2.3\mu{}$m to $0.51\mu{}$m luminosity ratio of $\log(L_{2.3}/L_{0.51})=-0.71$. One object out of nineteen corresponds to an inferred fraction for each of these special object classes (BAL, radio-loud, hot dust-poor) in the range $0.035-0.15$ (\change{beta distribution, 68\% confidence interval, see \sect{merger_fraction} below}). This is consistent with these fractions for the \zeq{2} luminous quasar population as a whole \citetext{$\simeq10-15$\% BALs \citealp{gibson_catalog_2009, allen_strong_2011}, $10-15$\% radio-loud \citealp{jiang_radioloud_2007}, $\simeq2$\% dust-poor \citealp{jun_physical_2013}}. \tab{sdss_data} also includes the measured $F160W$ magnitude of each quasar from our HST imaging (see next section), and the rest-frame $V$-band absolute magnitude, calculated using the \citet{vanden_berk_composite_2001} median quasar spectrum for the $k$-correction.

\begin{deluxetable\flexstar}{lrrrrrrrc}
\rotate
\tablewidth{0pt}
\tablecolumns{9}
\tablecaption{Properties of Observed Quasars}

\tablehead{
\colhead{Quasar} & 
\colhead{Redshift} & 
\colhead{$L_{3000}$} & 
\colhead{FWHM} &
\colhead{$M_\mathrm{BH}$} & 
\colhead{$L/L_\mathrm{Edd}$} & 
\colhead{$m_{F160W}$} &
\colhead{$M_V$} &
\colhead{Notes} \\
\colhead{(SDSS J)} & 
\colhead{$z$} & 
\colhead{$\log{}(\LSun{})$} & 
\colhead{\ion{Mg}{2} (km/s)} &
\colhead{$\log{}(\MSun{})$} & 
\colhead{} & 
\colhead{mag} &
\colhead{mag} &
\colhead{} \\
\colhead{(1)} &
\colhead{(2)} &
\colhead{(3)} &
\colhead{(4)} &
\colhead{(5)} &
\colhead{(6)} &
\colhead{(7)} &
\colhead{(8)} & 
\colhead{(9)}
}
\startdata
\begin{filecontents*}{tables/sdss_data.csv}
081518.99$+$103711.5, 2.021, -1.59, 12.72, 3600, 9.3, 9.3,  0.49, 18.55, -27.49, \nodata
082510.09$+$031801.4, 2.035, -1.19, 12.91, 4900, 9.4, 9.7,  0.28, 17.81, -28.27, \nodata
085117.41$+$301838.7, 1.917, -1.61, 12.54, 3000, 9.1, 9.0,  0.65, 18.85, -27.08, \nodata
094737.70$+$110843.3, 1.905, -1.34, 12.66, 6200, 9.4, 9.7,  0.11, 18.91, -27.01, \ion{C}{4} BAL
102719.13$+$584114.3, 2.020, -1.74, 12.67, 3700, 9.2, 9.3,  0.57, 18.67, -27.38, \nodata
113820.35$+$565652.8, 1.917, -1.26, 12.76, 4300, 9.6, 9.5,  0.30, 18.12, -27.81, \nodata
120305.42$+$481313.1, 1.988, -0.88, 12.79, 4100, 9.2, 9.4,  0.27, 18.18, -27.84, \nodata
123011.84$+$401442.9, 2.049, -1.18, 13.17, 5400, 9.9, 9.9,  0.28, 17.41, -28.67, \nodata
124949.65$+$593216.9, 2.052, -1.03, 12.88, 4100, 9.4, 9.5,  0.35, 18.11, -27.99, \nodata
131501.14$+$533314.1, 1.921, -1.75, 12.70, 8400, 9.8, 10.0, 0.10, 18.45, -27.47, \nodata
131535.42$+$253643.9, 1.926, -1.89, 12.57, 3200, 9.2, 9.1,  0.71, 18.50, -27.44, \nodata
135851.73$+$540805.3, 2.066, -1.71, 12.80, 3700, 9.5, 9.4,  0.54, 18.27, -27.83, \nodata
143645.80$+$633637.9, 2.066, -1.24, 13.22, 6800, 9.9, 10.1, 0.22, 17.06, -29.04, Radio-loud
145645.53$+$110142.6, 2.017, -1.36, 12.79, 3200, 9.4, 9.2,  0.65, 18.20, -27.86, \nodata
155447.85$+$194502.7, 2.091, -1.05, 12.71, 5900, 9.5, 9.7,  0.14, 18.77, -27.37, \nodata
215006.72$+$120620.6, 1.993, -1.75, 12.66, 5100, 9.6, 9.5,  0.29, 18.55, -27.46, \nodata
215954.45$-$002150.1, 1.963, -1.46, 13.33, 4000, 9.7, 9.8,  0.74, 16.87, -29.12, Hot dust-poor
220811.62$-$083235.1, 1.923, -1.85, 12.77, 3300, 9.3, 9.2,  0.63, 18.50, -27.43, \nodata
232300.06$+$151002.4, 1.989, -1.06, 12.81, 5400, 9.5, 9.7,  0.19, 18.13, -27.89, \nodata 
\end{filecontents*}
\csvreader[no table,no head]{tables/sdss_data.csv}{}%
{\csvcoli & \csvcolii & \csvcoliv & \csvcolv & \csvcolvii & \csvcolviii & \csvcolix & $\csvcolx$ & \csvcolxi \\}
\enddata
\tablecomments{Properties of \zeq{2} quasars. Columns 1--5 are from the catalog of \citet{shen_catalog_2011}. Column 1: SDSS name, including the full sexagesimal celestial coordinates; Column 2: Systematic redshift; Column 3: Rest-frame 3000\AA{} luminosity; Column 4: FWHM of the broad component of the \ion{Mg}{2} line emission (median fractional error is 11\%); Column 5: \ion{Mg}{2} virial BH mass \citep[calibration of][]{shen_catalog_2011}; Column 6: Eddington ratio, derived using the \ion{Mg}{2} BH masses and the catalog bolometric luminosity; Column 7: $F160W$ observed magnitude from our HST observations (photometric error is $\simeq 0.05$~mag); Column 8: $V$-band absolute magnitude ($k$-corrected using the \citet{vanden_berk_composite_2001} median quasar spectrum); Column 9: Special notes or features
}
\label{tab:sdss_data}
\end{deluxetable\flexstar}

\subsection{Hubble Space Telescope Data}
\label{sec:hst_data}
All 19 quasars were observed with the WFC3 infrared channel using the $F160W$ filter (broad H-band). None of the quasars had any existing HST imaging at rest-frame optical wavelengths. As a SNAP program, the observations by necessity have short integration times of 1597 seconds per target ($<1$ orbit). Each observation was split into four exposures, dithered using the standard four-point half-pixel box pattern to improve the PSF sampling, and to assist in the rejection of bad pixels and cosmic rays.

Our data processing began with individual flat-fielded, flux-calibrated exposures delivered by the HST archive. The four exposures for each pointing were combined using the \soft{astrodrizzle} software package \citep{koekemoer_multidrizzle:_2002, koekemoer_2012_2013} with an output plate scale of 0\farcs{}060 per pixel and a \texttt{pixfrac} parameter of 0.8. For $F160W$ observations, this samples the PSF with 2 pixels per FWHM and provides relatively uniform weighting of the individual pixels. We used ``ERR'' (minimum variance estimator) weighting for the final image combination step. We also generated variance maps that include all sources of noise, including uncertainty from the \soft{calwf3} count rate determination, by copying the WFC3 ``ERR'' arrays into the standard image arrays, and re-running the drizzle process using the same parameters and weighting scheme. The variance maps are a requirement for our analysis since the count rate from the quasar point source is significantly higher than that of the sky, and under-estimated errors can lead to problems with multi-component fitting, as described in \sect{bayesian_model} below.

Despite the short exposure time, the excellent sensitivity of WFC3 IR and low on-orbit near-infrared sky allow the images to reach $1\,\sigma{}$ limiting surface brightnesses in the range $\simeq24.0-24.5$~\magarc{}. This is sufficient to detect tidal features of major mergers between luminous galaxies, as illustrated by the WFC3 imaging of the most luminous, distorted \zsim{2} merger in the \change{GOODS/ERS2} field \citep{van_dokkum_hubble_2010, ferreras_road_2012}. We note that the true achieved surface brightness sensitivity is a function of distance from the quasar point source (due to shot noise from the removed point source). \change{The sensitivity to low surface brightness features is discussed in detail in Appendix~\ref{sec:appendix_images}.}

\subsection{Potential Selection Biases}
\label{sec:biases}
\change{
As mentioned in \sect{intro}, particular care must be taken in selecting quasar samples since there is no unbiased selection method that captures the entire population. Further, selecting non-random subsamples from even large survey catalogs may introduce additional biases, and observational constraints may bias sensitivity to features of interest. For our study, two kinds of bias are the most salient. First, if a universal evolutionary sequence such as that proposed by \citet{sanders_ultraluminous_1988} exists, certain quasar selection methods may bias toward a particular phase within that evolution. Second, since we are interested in assessing evidence for mergers via morphological signatures, observational biases affecting sensitivity to those signatures are relevant, in particular surface brightness sensitivity and the rest-frame emitted wavelength. We discuss the observational biases of this program in detail in Appendix~\ref{sec:appendix_images}.

Selecting optically luminous broad-line sources (Type~1 quasars) by definition selects objects where the central accretion disk is essentially unobscured along the line of sight. Highly obscured or very red (Type~2) sources missed by this selection may include objects similar to the unobscured sources but viewed from different angles, as well as fundamentally different objects that contain systematically more dust or different dust geometries. For luminous quasars such as those in this study ($L_{Bol}\gtrsim{}10^{47}$~erg~s$^{-1}$), the obscured fraction due to anisotropy of the dust torus is estimated from Type~1 quasar SED modeling to be $\simeq{}20-50$\% \citep{lusso_obscured_2013}. Mid-infrared quasar selection methods directly estimate the \emph{total} luminous Type~2 fraction --- \ie{} the above torus-obscured quasars but also those obscured due to fundamentally different dust geometries --- to be $\lesssim{}50$\% \citep{donley_spitzers_2008, assef_half_2015}. We thus expect $\lesssim{}30$\% of luminous quasars to be hosted in systems with much larger dust covering fractions than our sample, whether these represent the putative ``buried'' evolution phase with the quasar completely enshrouded in dust, or simply a sub-population of luminous quasars hosted in ULIRG-like dust-rich galaxies. This also sets an upper limit on a buried phase duration of at most 30\% of the average quasar lifetime. We argue in \sect{merger_preference} below that this timescale is insufficient for dynamical effects to erase the major merger signatures we are searching for.

Besides the implicit luminosity and unobscured line-of-sight constraints, there are potential biases from explicitly selecting quasars with high-mass black holes. In particular, since the (active) black hole mass function drops sharply at high masses, there could be concern that such high-mass quasars are preferentially near the end of their accretion phases. Although the volume density of active black holes drops at high mass, the Eddington ratio distribution function does not depend strongly on black hole mass \citep{schulze_cosmic_2015}, so even high-mass black holes like those in our sample seem to have accretion episodes statistically similar to lower-mass black holes. We discuss merger signatures as a function of black hole mass in \sect{merger_preference}, but note here that even if the quasar duty cycle is near unity and these quasars grow continuously near the sample average $\langle{}L/L_{Edd}\rangle{}=0.40$, fifteen of the nineteen could continue growing for $>10^{8}$ years before they exceeded the maximum mass of our selection region, on the long side for quasar lifetime estimates \citep{martini_quasar_2001, yu_observational_2002, shen_clustering_2007}. Thus, although these high-mass black holes might be experiencing their last major growth phase, there is no a priori reason to assume we are observing the quasars late in the \emph{current} active phase.
}

\section{Point Source Subtraction}
\label{sec:psf_sub}
\subsection{Point Spread Function Models}
\label{sec:psf_model}
As a SNAP program, our HST data did not have dedicated observations of stars to measure the instrument and telescope PSF. The focus of all HST instruments is affected by the telescope's thermal environment, with changes in solar illumination resulting in de-space of the secondary mirror, the so-called ``spacecraft breathing'' effect \citep{bely_orbital_1993, hershey_modelling_1998}. For applications requiring precise PSF matching, the exposure focal history is best matched by extracting PSF stars from the same images. However, the WFC3 infrared channel PSF has other aberrations (\eg{} coma, astigmatism) that vary with position within the WFC3 field of view. Simulated WFC3 PSF models are currently poor matches to observations \citep[\eg{}][]{mechtley_near-infrared_2012, biretta_tinytim_2012}, so we chose to build a library of empirical PSF models from WFC3 archival data of high S/N stars near the center of the field of view, where all the quasar targets were observed.

To find similar exposures from which to extract PSF models, we searched the HST archive for all single-orbit $F160W$ observations using a 4-exposure dither pattern. We then identified all point sources falling within 0\farcm{}5 of the WFC3 field of view center, and excluded known quasars or radio sources in the NASA/IPAC Extragalactic Database or were from HST programs specifically targeting AGN. We visually inspected the remaining PSF stars, and excluded any that were contaminated by background galaxies or whose flux distribution was significantly elliptical, \ie{} were likely stellar binaries. This left us with a library of 8 stars\footnote{We note that 5 of these 8 were dedicated PSF star observations from two \emph{other} quasar host programs: GO 12332 (PI: Windhorst) and GO 12974 (PI: Mechtley).} with high S/N, but which still had an accurate count rate determination in their cores. We created drizzled images and variance maps of these stars using the same plate scale and weighting scheme described in \sect{hst_data} above.

While a posteriori estimation of HST focus is possible from on-orbit thermal measurements using the HST Focus Model \citep{cox_evaluation_2011}, there is yet another (essentially) degenerate source of PSF mismatch, namely the object SED through the broadband $F160W$ filter. \citet{bahcall_hubble_1997} noted that for diffraction-limited HST observations through a broadband filter, the color of an observed star (effectively, the SED-weighted average of the monochromatic PSFs) can significantly affect the quality of quasar point source subtractions if the quasar and star SEDs are poorly matched. To first order, redder objects will have a slightly broader PSF (\eg{} as measured by FWHM), and bluer objects slightly narrower. Since we do not have detailed information about the shape of the quasar or star SEDs through the filter, and because of the degeneracy with focus, we simply leave the choice of PSF as a free parameter during the fitting procedure, which also allows matching of higher-order features (\eg{} differences in diffraction spike patterns).


\subsection{Bayesian Modeling Method}
\label{sec:bayesian_model}
In the presence of detectable host galaxy flux, fitting only a single point source when attempting to remove the central quasar light tends to over-subtract the point source, especially for bright or centrally concentrated host galaxies. We therefore adopted a simultaneous fitting technique that models the point source and the underlying host galaxy flux distribution, approximating the latter with a S\'ersic profile \citep[following][]{mechtley_near-infrared_2012}. We stress that the actual morphologies of the host galaxies may differ from symmetric ellipsoids; the purpose of the S\'ersic model is simply to provide a first-order approximation of the surface brightness distribution with a flexible parameterization.

We first attempted the two-component fit using the software \soft{Galfit} \citep{peng_detailed_2002, peng_detailed_2010}, currently the most widely-used 2D surface brightness modeling software.  However, this software employs several design decisions that make it less desirable for this particular problem \citep[\eg{} as noted in \S{}6.2 of][]{peng_detailed_2010}. 

First, \soft{Galfit} uses a standard Levenberg-Marquardt gradient descent method to perform least-squares minimization when fitting models. This method involves calculating a gradient image during each iteration to determine the parameter values to use for the subsequent iteration. Extremely compact models --- \eg{} a S\'ersic profile with small effective radius ($r_e$) and large index ($n$) --- have most of their gradient information contained within a single pixel, and so the S\'ersic degrees of freedom are prone to fitting aberrant pixels (\eg{} from PSF mismatch). This essentially creates a false minimum in parameter space, from which the gradient descent cannot escape.

A second, related problem is that \soft{Galfit} assumes that the supplied PSF is without error. Even without systematic PSF uncertainties --- \ie{} a PSF exactly matching the telescope focal history, spectral energy distribution of the quasar point source, etc. --- the photon noise can be large enough to cause problems. While our PSF stars were selected to have higher S/N than the quasars, they exceed the quasar S/N by less than a factor of 10. This means that when performing the PSF subtraction, the PSF can contribute up to $\simeq$30\% of the per-pixel RMS error.

Finally, the model is expected to have covariant parameters, such as the relative flux normalizations of the point source component and S\'ersic component. These covariances are not quantified by \soft{Galfit}.

To address these problems, we developed our own Markov Chain Monte Carlo (MCMC) simultaneous fitting software, \soft{psfMC} \citep{mechtley_markov_2014}\footnote{The details of the software implementation are given in \citet{mechtley_markov_2014}. The software, documentation, examples, and source code are available at: \url{https://github.com/mmechtley/psfMC}}. As an MCMC parameter estimator, it addresses the first and third problems intrinsically, by allowing the user to provide prior probability distributions for the model parameters, and providing as an output product the full posterior probability distribution of model parameters given the observed data. The second problem is addressed by propagating a supplied PSF variance map during the convolution process.

Our software is built upon the \soft{pyMC} Python module for Bayesian stochastic modeling \citep{patil_pymc:_2010}. The computational book-keeping tasks --- such as implementing the MCMC sampler, setting proposal distributions (see below), evaluating prior probabilities for a given set of parameters, and saving sample traces to disk --- are handled by \soft{pyMC}. We provide a framework that allows the user to simultaneously model an arbitrary number of model components (at this time, point sources, S\'ersic profiles, and sky background). The free parameters for each component (\eg{} position, total magnitude, S\'ersic index, etc.) can either be supplied as a fixed numeric value or as an arbitrary prior probability distribution. An arbitrary number of samples can be drawn from the posterior distribution.

Samples are drawn from the posterior distribution using the Metropolis-Hastings algorithm \citep{metropolis_equation_1953, hastings_monte_1970}, which accepts a proposed sample probabilistically based on the ratio of its posterior probability to that of the previous sample in the Markov Chain. Each time a proposed sample is drawn from the parameter space, \soft{psfMC} generates a model image of the intrinsic surface brightness distribution described by the parameters (hereafter, ``raw model''). This raw model is then used to generate two further images --- one convolved with the PSF (``convolved model''), and a model variance map.

The model variance map is simply the square of the raw model image convolved with the PSF variance map. The intensity, $I_C(p)$, of a pixel $p$ in the convolved model is given in \eqn{conv_int} below, where the summation is over all pixels $q$. $I_R(q)$ is the raw model intensity, and $K(p,q)$ is the PSF weight for pixel $p$ with the kernel centered at pixel $q$. If the kernel weights have associated variances $\sigma{}^2_K(p,q)$, then the variance propagates in the usual way, and the convolved variance of pixel $p$ is given by \eqn{conv_var} below. This assumes that the off-diagonal covariance terms are zero (the noise between pixels is not correlated), which is almost true for the drizzle parameters used, and is a standard assumption in 2D surface brightness modeling.

\begin{equation} \label{eqn:conv_int}
I_C(p) = \sum_{q} I_R(q) K(p,q)
\end{equation}
\begin{equation} \label{eqn:conv_var}
\sigma^2_C(p) = \sum_{q} I_R(q)^2 \sigma^2_K(p,q)
\end{equation}

The conditional probability of each observed pixel value given the model is then calculated, using a normal distribution with mean equal to the PSF-convolved model, and a variance equal to the summed convolved model and observed variances. The likelihood function is then the joint probability of these individual pixel probabilities, which is then multiplied by the prior probabilities of the parameter values to generate the sample's posterior probability. The sample is then accepted or rejected based on the Metropolis criterion.

\subsection{Model Parameter Estimation and Convergence Checking}
\label{sec:model_comps}
Each quasar in our sample was modeled by two simultaneously-fit components --- a point source and a S\'ersic profile. Although the software enables more complex multi-component fitting, the number of required samples (and thus computation time) increases steeply with additional parameters. Since we needed to model a total of 179 images (19 quasar images and 160 comparison galaxy images, see \sect{comparison_selection}), this argued against iterative hand-crafting of more complex models for hosts that show multiple nuclei or other more complicated structures. Instead, we masked out other galaxies in each image with flux peaks that were separate from central source (effectively, any galaxy with $\gtrsim{}1\farcs{}5$ separation), excluding them from the fit. This ensures that the S\'ersic profile free parameters are used to fit the light distribution of the host galaxy (or galaxies) surrounding the quasar, rather than neighboring or foreground galaxies. This 2-component model is a computational compromise, one S\'ersic profile being more appropriate than none to avoid over-subtracting the point source, so we caution against over-interpreting the model parameter estimates. For instance, positional offsets between the point sources and the S\'ersic profile centers should be interpreted as evidence for asymmetric flux distributions or multiple components in the host, rather than physical separation between the black hole and the center of its host.

The prior distributions adopted for the model parameters are summarized in \tab{mcmc_priors}. At \zsim{2}, the drizzled 0\farcs{}060 linear pixel scale corresponds to a physical size of $0.52$~kpc. The ranges on the parameter priors were selected to model the entire range of values that are both physically reasonable and detectable. In particular, the Weibull distribution is used to approximate the observed distribution of S\'ersic indexes \citep{yoon_new_2011, ryan_jr._size_2012}\footnote{The Weibull distribution is a continuous, asymmetric distribution with non-negative support. Its probability density function is given by: $P(x; \alpha, \beta) = \frac{\alpha}{\beta} (\frac{x}{\beta})^{\alpha-1}\exp[-(x/\beta)^\alpha]$}.

Four chains were run for each model, each 100,000 samples with the first 50,000 discarded as a burn-in period, allowing the chains an opportunity to converge in parameter space before they are retained for analysis. Each chain is fully independent, with no parallel tempering of proposal distributions used. We use the Gelman-Rubin Potential Scale Reduction Factor to assess convergence \citep[$\widehat{R}$,][]{gelman_inference_1992}. This summary statistic compares the variance within individual chains to the variance among the chains to estimate how much sharper the posterior distribution for a parameter could be made by running for more samples. To consider a parameter's posterior estimate converged, we require $\widehat{R} < 1.05$, \ie{} the estimated potential reduction in the scale of a parameter's univariate marginalized posterior distribution is less than 5\%. Samplers whose chains had not converged (one or more parameters had $\widehat{R} \geq{} 1.05$) after 100,000 samples were continued for another 50,000 samples to increase the burn-in time, up to two additional times. Objects requiring longer convergence times are generally those where the host galaxies are only marginally detected. In all cases the point source parameters (position and magnitude) were well-determined and met the above convergence criterion. In four of nineteen models, one or more S\'ersic parameters still had $\widehat{R} \geq{} 1.05$. Since the goal of the modeling is point source subtraction, and single S\'ersic models are a contrived simplification of the true flux distributions, we consider these converged for the purpose of our experiment.

\begin{deluxetable}{lll}
\tablecolumns{3}
\tablewidth{0pt}
\tablecaption{Adopted Prior Distributions of Model Parameters}
\tablehead{
\colhead{Model Parameter} & 
\colhead{Distribution} & 
\colhead{Value Range}
}
\startdata
\multicolumn{3}{l}{Whole-Image Parameters} \\
\tableline \\[-0.8em]
Stellar PSF Image         & Discrete Uniform  & $1-8$ (list index)\\
Sky Background            & Normal            & $0\pm0.01$~e$^-$s$^{-1}$\\
\\[-0.5em] \multicolumn{3}{l}{Point Source Component} \\
\tableline \\[-0.8em]
X,Y Position              & Normal            & Centroid $\pm{}4$~pix \\
Total Magnitude           & Uniform           & $m_H \,^{-0.2}_{+1.5}$~mag \\
\\[-0.5em] \multicolumn{3}{l}{S\'ersic Component} \\
\tableline \\[-0.8em]
X,Y Position              & Normal            & Centroid $\pm{}8$~pix \\
Total Magnitude           & Uniform           & $m_H-26$~mag\\
Eff. Radius (Major)  & Uniform           & $1.0-15.0$~pix\\
Eff. Radius (Minor)  & Uniform           & $1.0-15.0$~pix\\
Index $n$                 & Weibull           & $\alpha{}=1.5, \beta{}=4$ \\
Position Angle            & Circular Uniform  & $0\degr-180\degr$ 
\enddata
\tablecomments{Ranges are for intrinsic quantities, before convolution with the PSF. ``Centroid'' and $m_H$ refer to the flux centroid and total $F160W$ magnitude of the (point-source dominated) quasar+host galaxy in the WFC3 image. Images are sky-subtracted during drizzling, but with some uncertainty, so the sky value is left as a free parameter.}
\label{tab:mcmc_priors}
\end{deluxetable}

\subsection{Posterior Model Analysis}
\label{sec:posterior_analysis}
The result of the MCMC fitting  process is a collection of samples representing the posterior probability distribution of the free parameters in \tab{mcmc_priors}, given the observed data. The \soft{psfMC} software uses this to create model images --- both before and after convolution with the PSF --- and an image with all point source components subtracted. These images can either be made from the single maximum a posteriori (MAP) sample, or posterior-weighted. Posterior-weighted here means that an average image is made from the model images of all the retained MCMC samples (hence, weighted by the posterior), rather than the single image from the MAP sample. The purpose of this is primarily diagnostic --- for models with well-constrained parameter estimates, the posterior-weighted and MAP images are almost identical. For models with poorly-constrained parameters or multiple posterior modes (\eg{} SDSS~J135851.73$+$540805.3), these will be visible in the posterior-weighted images, but not the MAP images, since MAP represent parameter estimates from only a single sample, rather than the full distribution of parameter values. \fig{quasar_resids} shows the observed quasar, the posterior-weighted model image before PSF convolution, and the image of the host galaxy after subtracting only the point source component of the model from the observed data, both with original sampling and smoothed by a $2\times2$ pixel gaussian.

\begin{figure\flexstar}
\centering
\includegraphics[width=\textwidth]{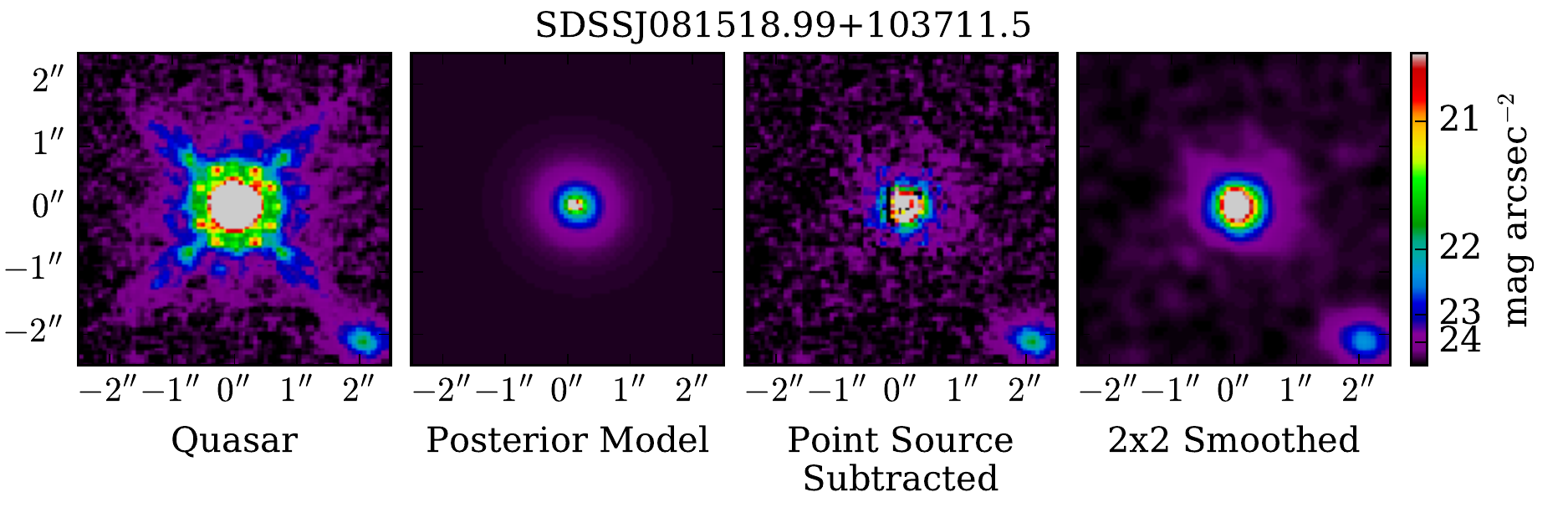}
\includegraphics[width=\textwidth]{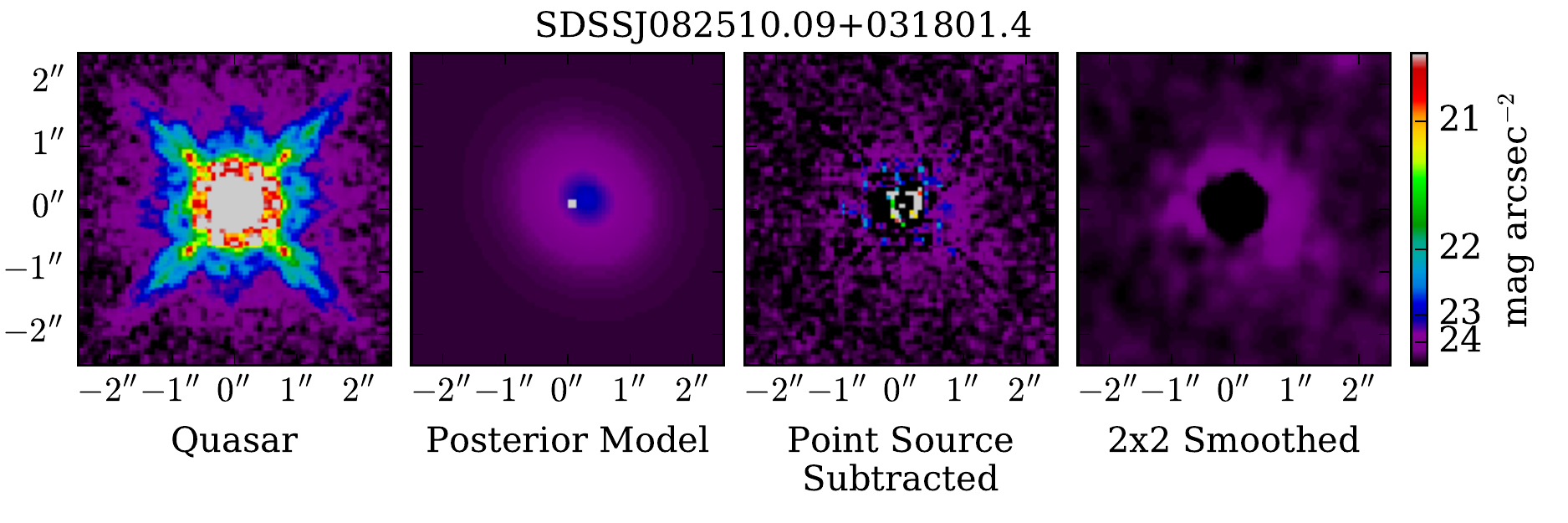}
\includegraphics[width=\textwidth]{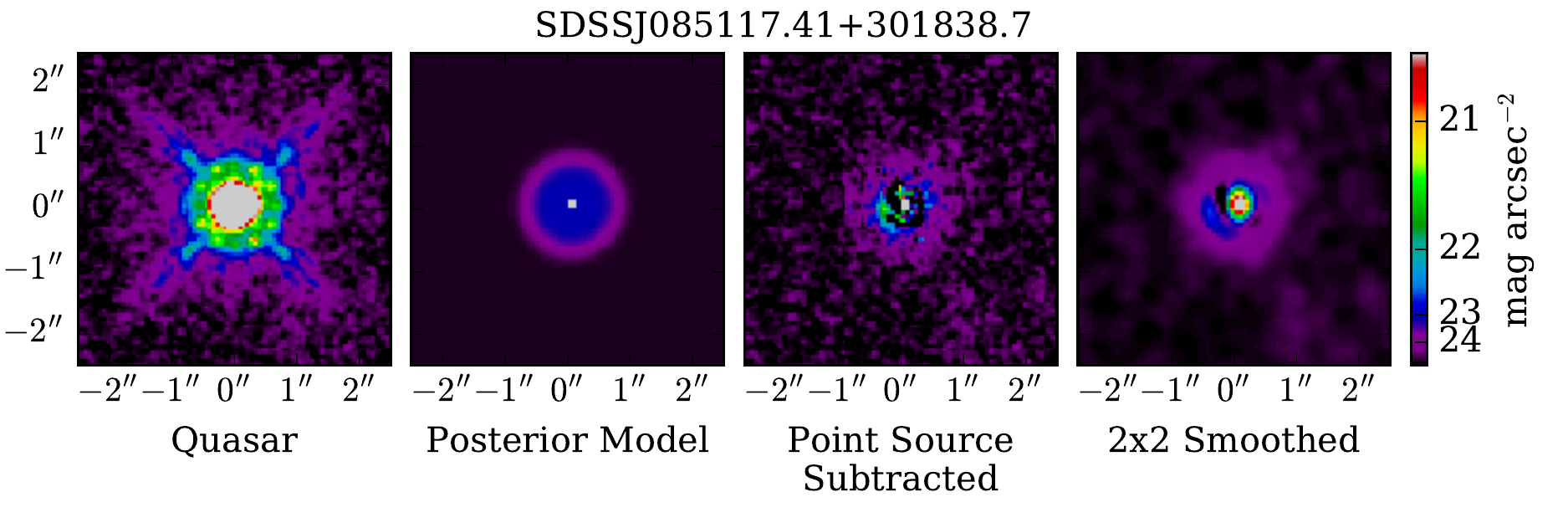}
\caption{Posterior-weighted model images for each quasar (see text for explanation). All images are displayed with the same arcsinh color stretch. Far left: Drizzled, undistorted WFC3 $F160W$ image with 0\farcs{}060 pixels. Middle left: Posterior-weighted model from MCMC fitting process, before convolution with the PSF. Middle right: Residual after subtracting only the posterior-weighted point source from original image. Far right: Same as middle right panel, but smoothed with a $\sigma=2$ pixel gaussian, to suppress high-frequency noise for the eye.
}
\label{fig:quasar_resids}
\end{figure\flexstar}

\begin{figure\flexstar}
\ContinuedFloat
\centering
\includegraphics[width=\textwidth]{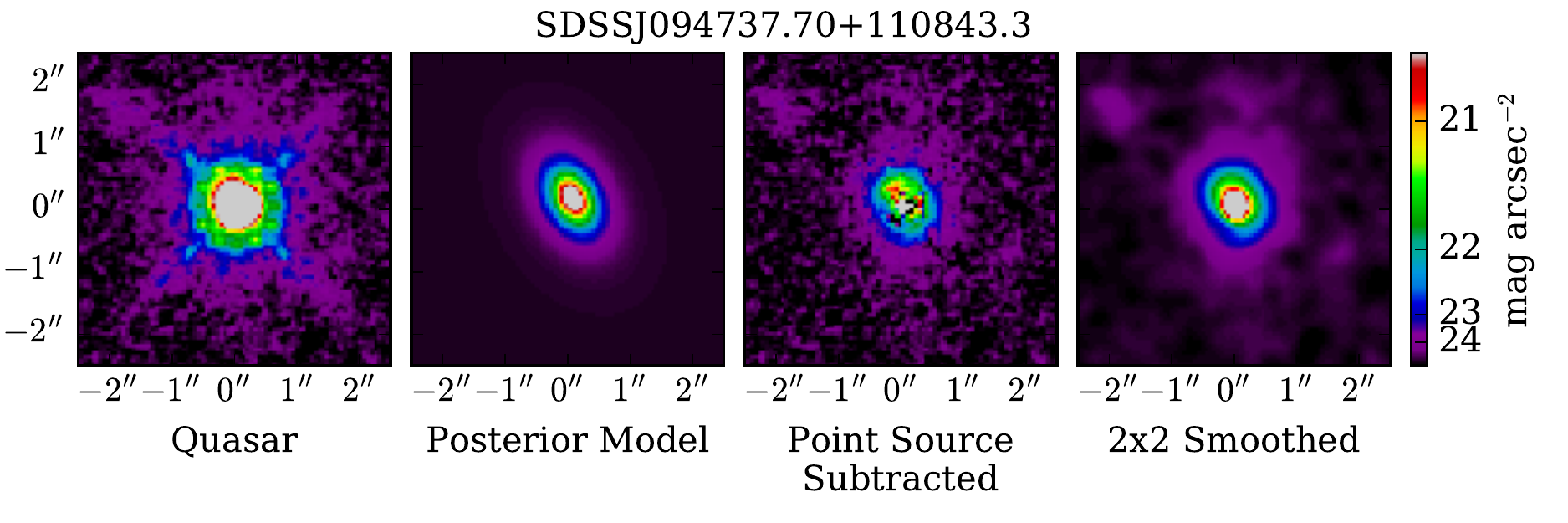}
\includegraphics[width=\textwidth]{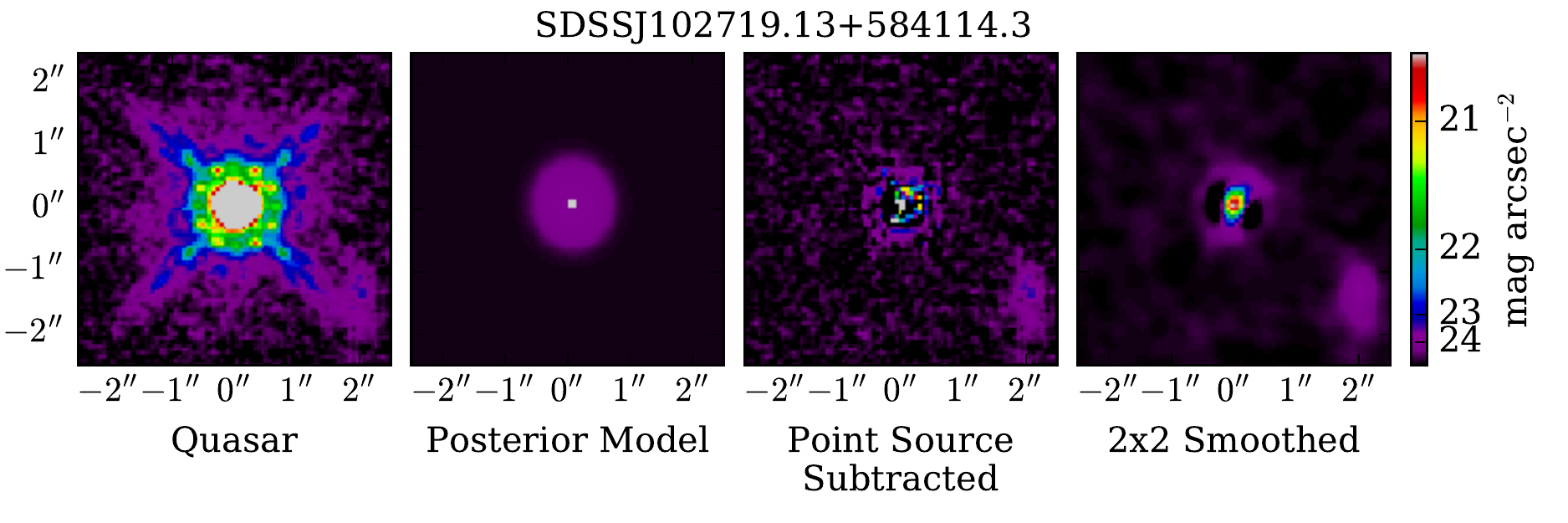}
\includegraphics[width=\textwidth]{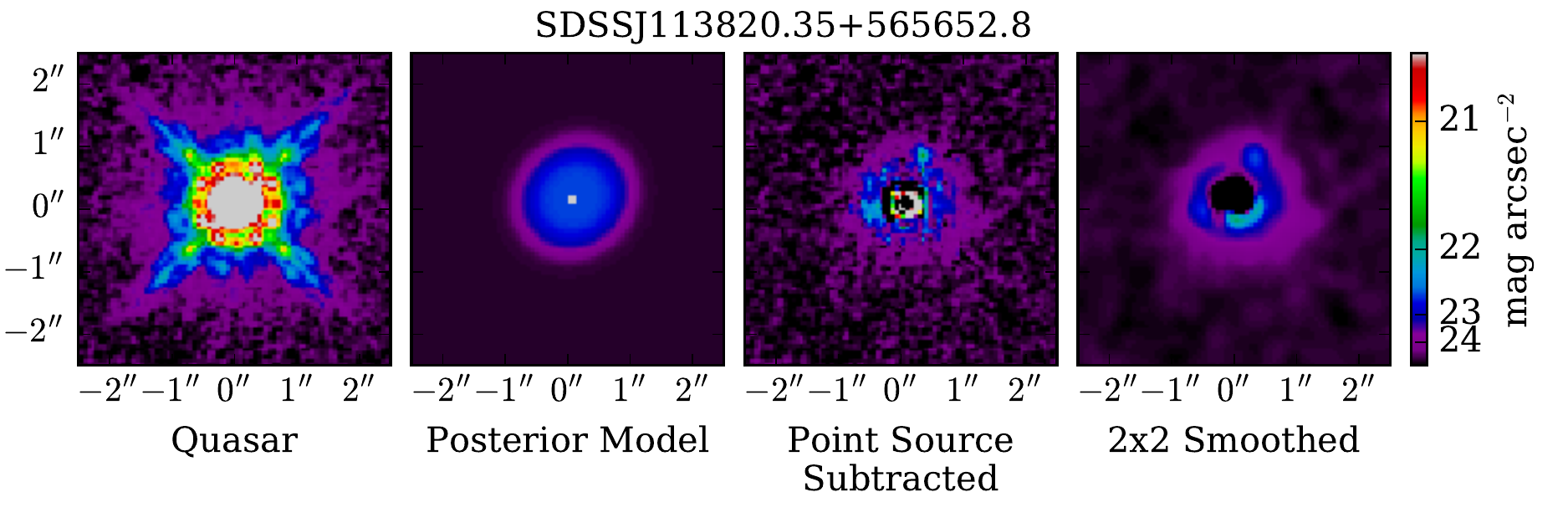}
\includegraphics[width=\textwidth]{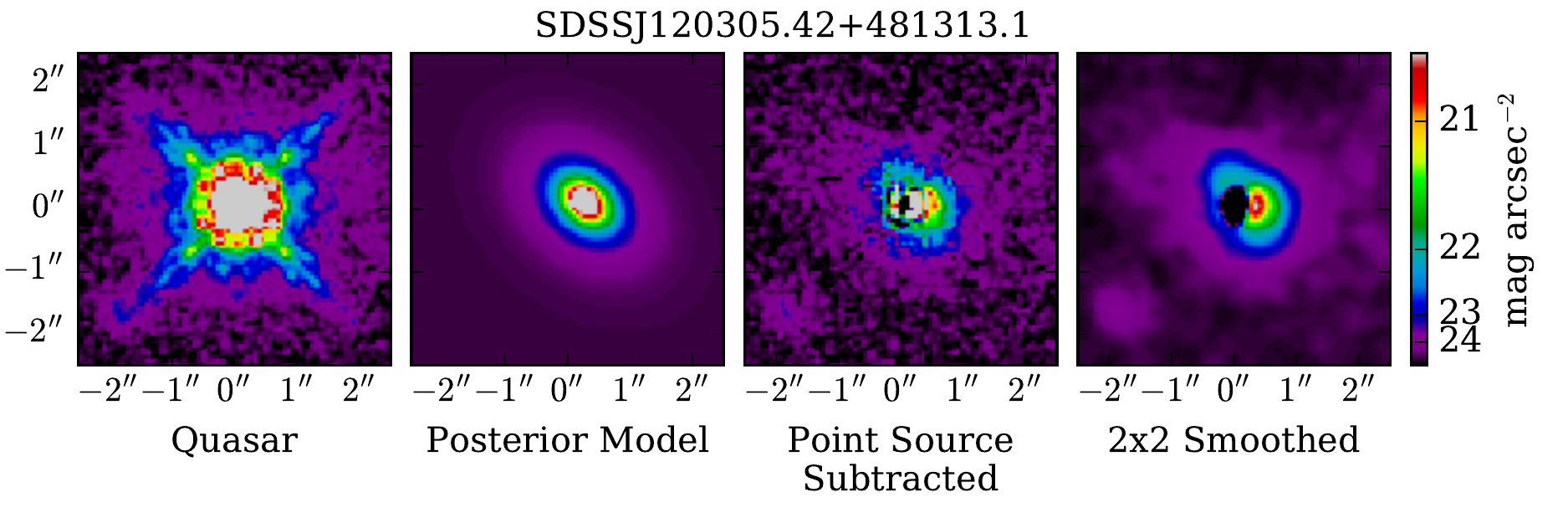}
\caption{(continued) Posterior-weighted models}
\end{figure\flexstar}

\begin{figure\flexstar}
\ContinuedFloat
\centering
\includegraphics[width=\textwidth]{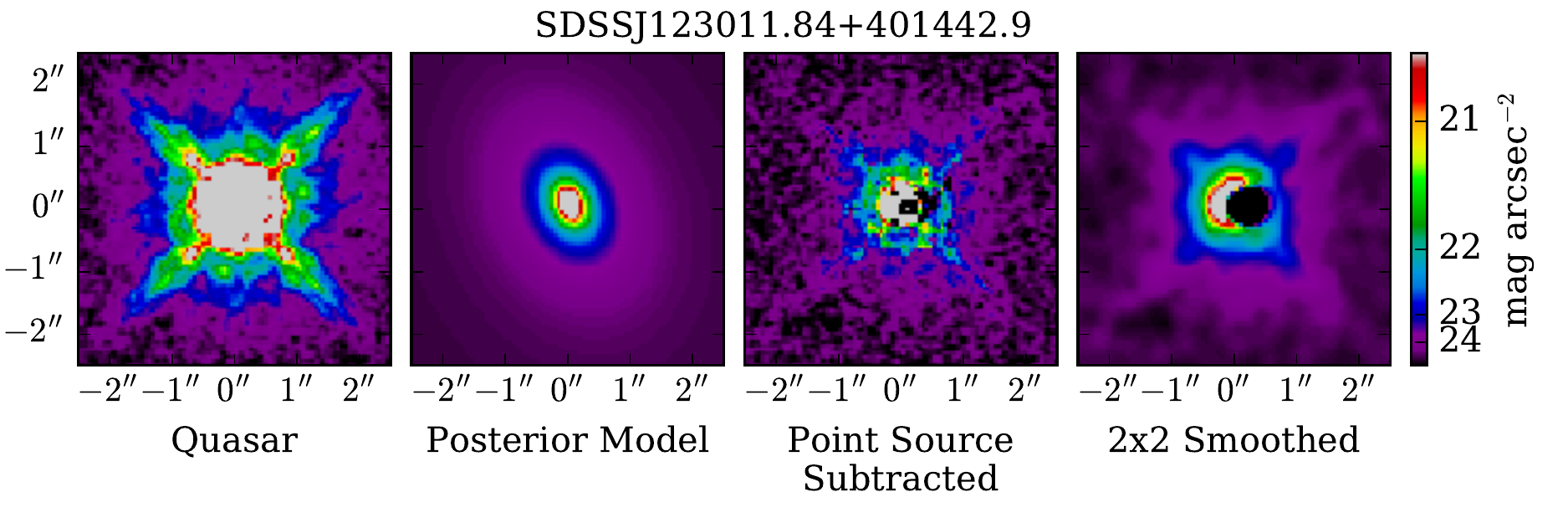}
\includegraphics[width=\textwidth]{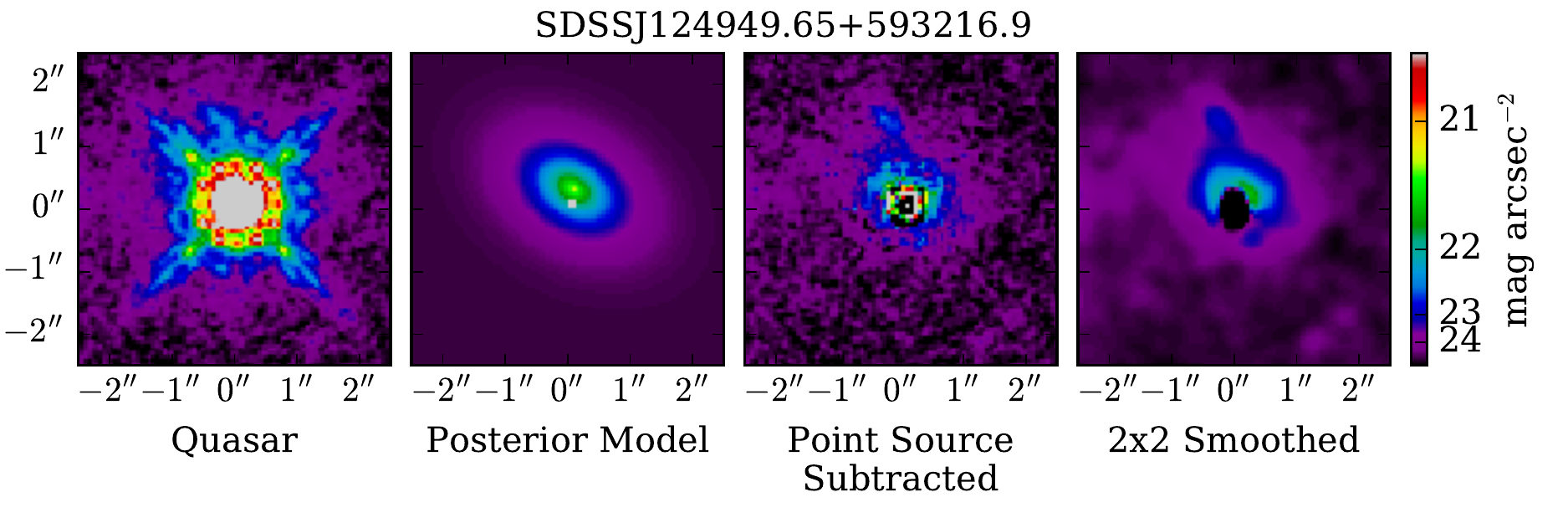}
\includegraphics[width=\textwidth]{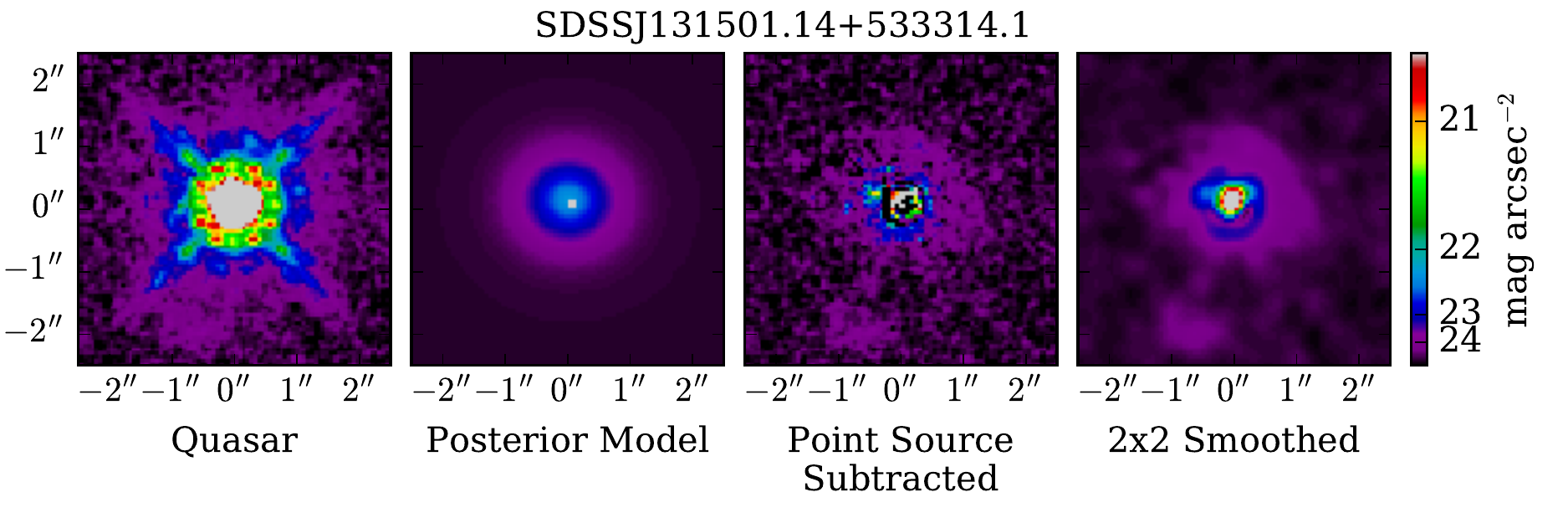}
\includegraphics[width=\textwidth]{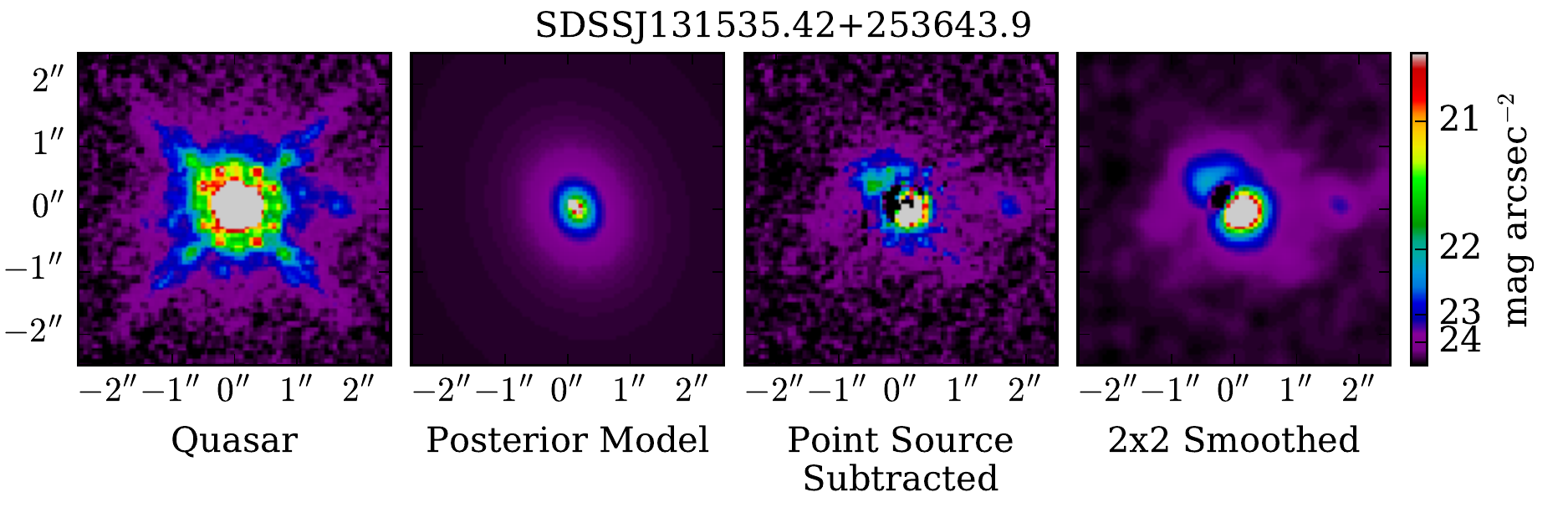}
\caption{(continued) Posterior-weighted models}
\end{figure\flexstar}

\begin{figure\flexstar}
\ContinuedFloat
\centering
\includegraphics[width=\textwidth]{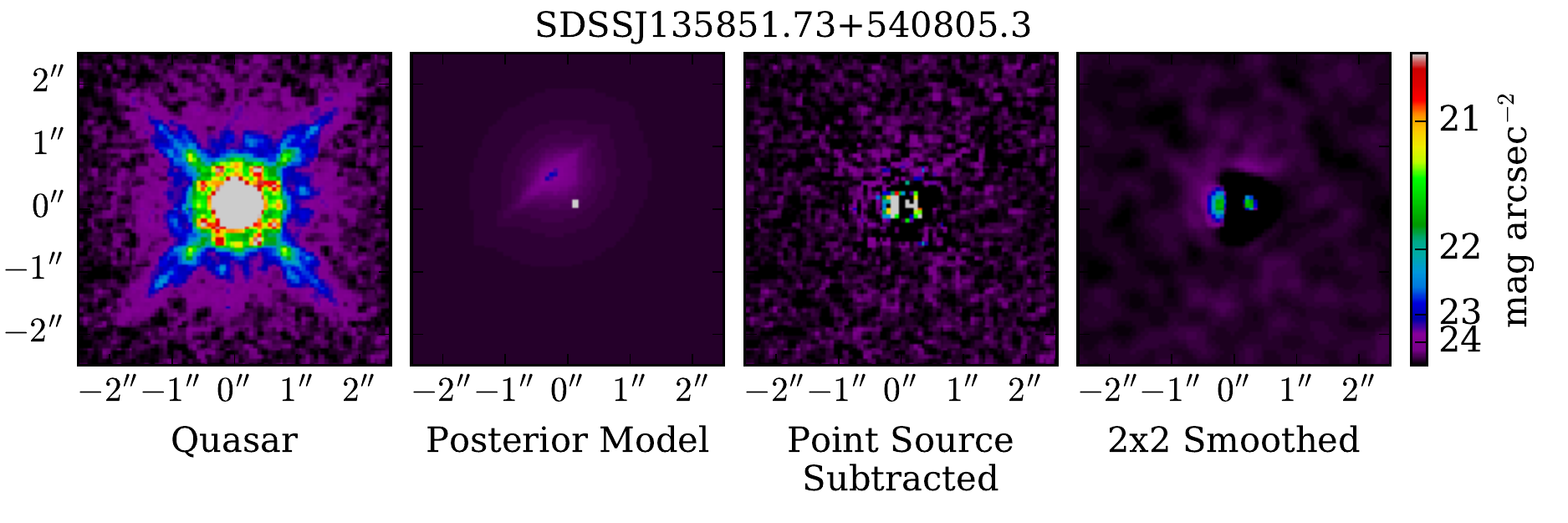}
\includegraphics[width=\textwidth]{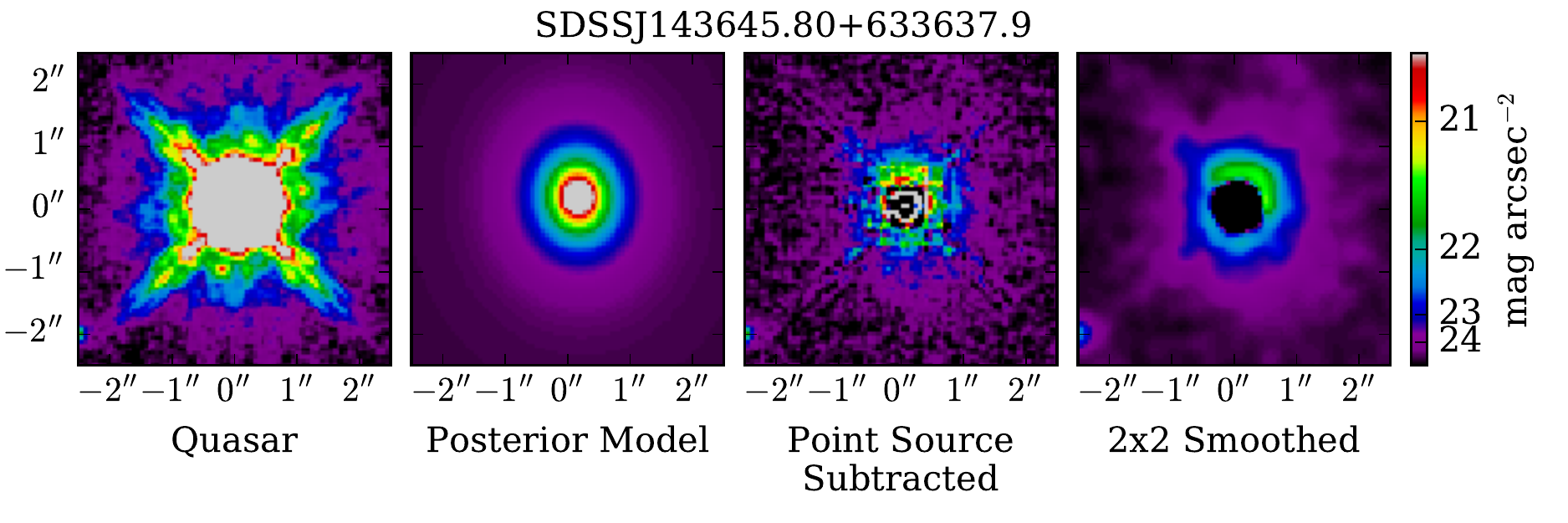}
\includegraphics[width=\textwidth]{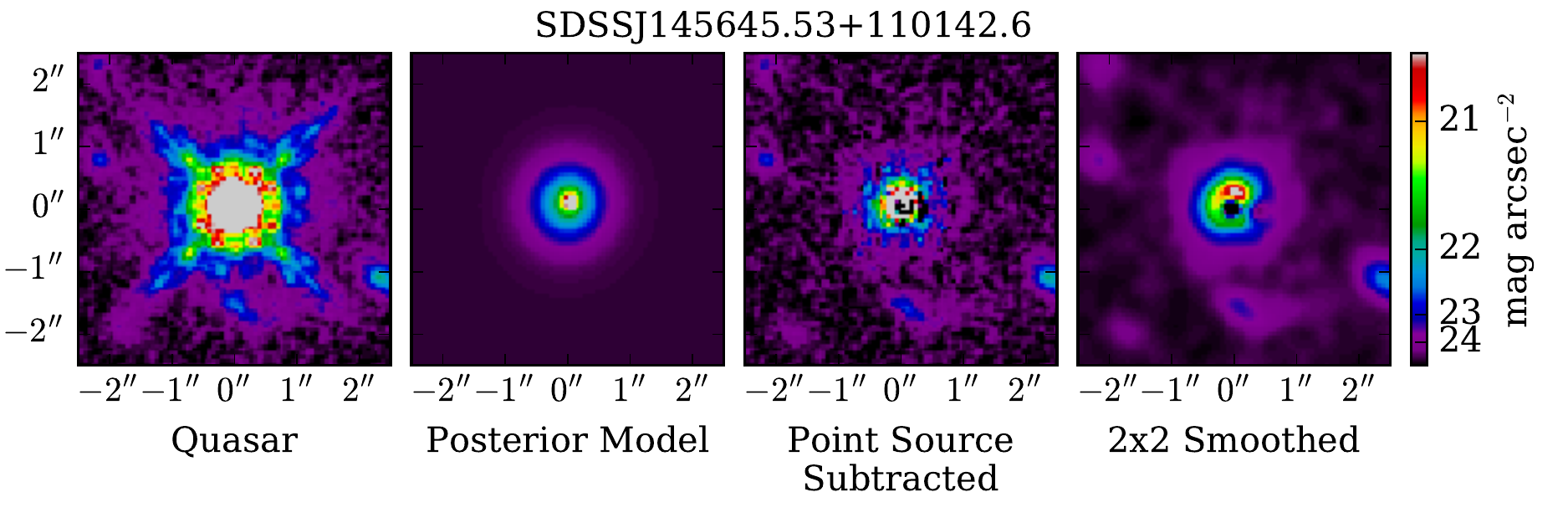}
\includegraphics[width=\textwidth]{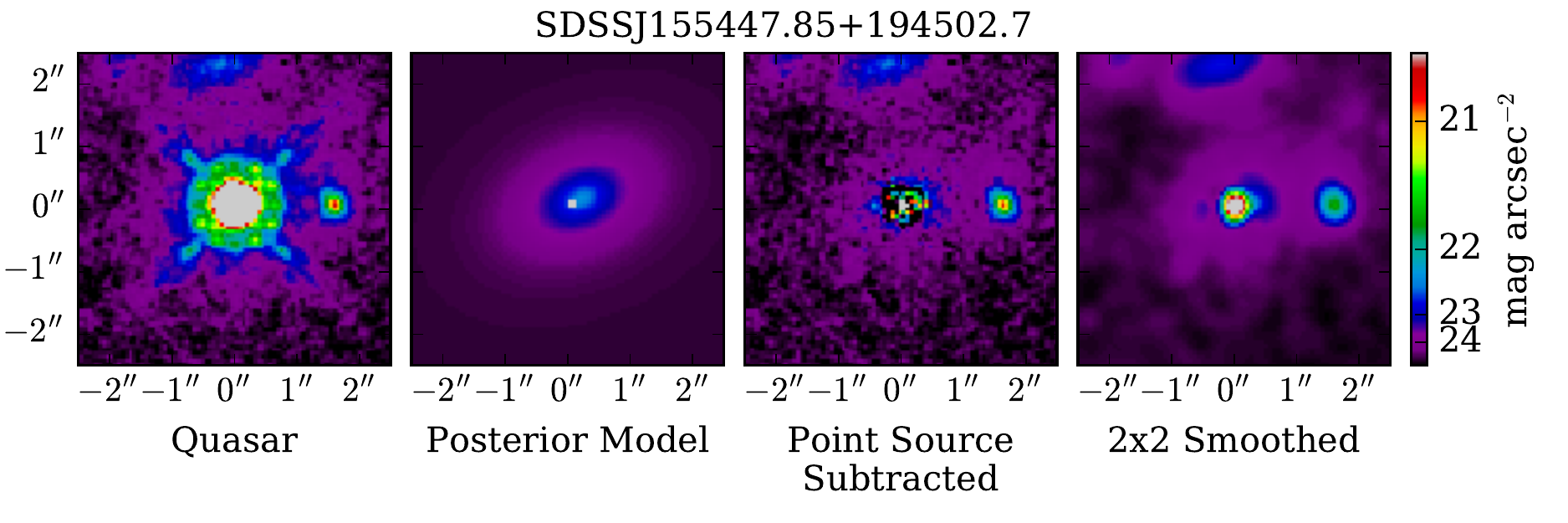}
\caption{(continued) Posterior-weighted models}
\end{figure\flexstar}

\begin{figure\flexstar}
\ContinuedFloat
\centering
\includegraphics[width=\textwidth]{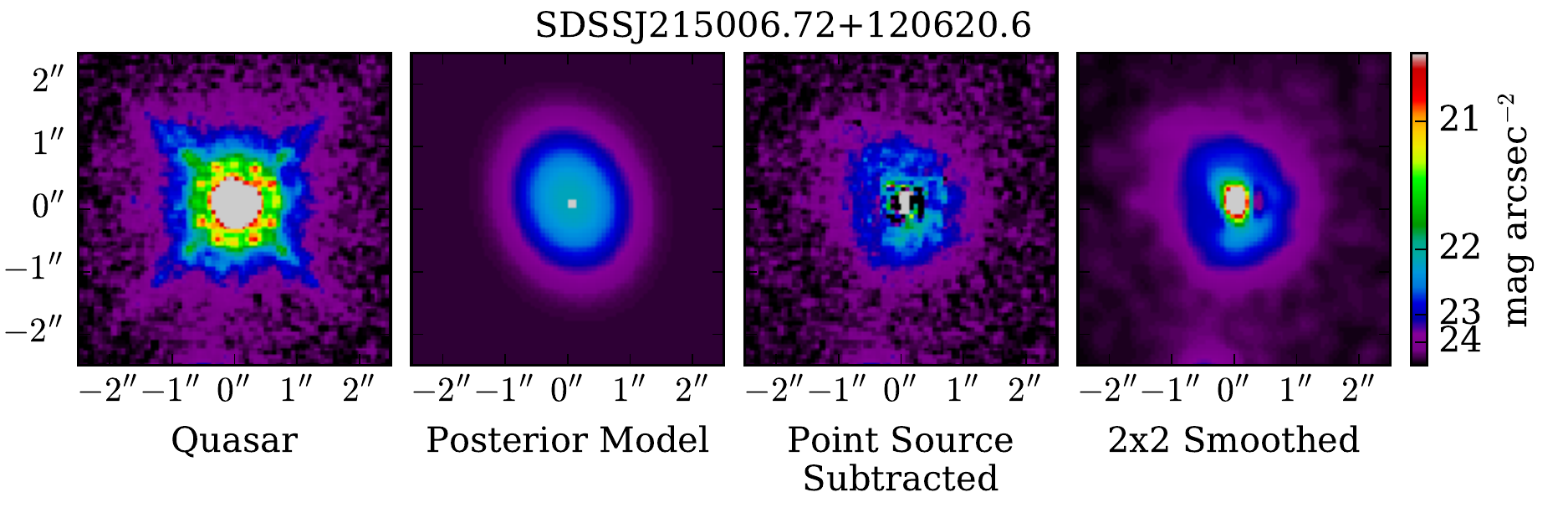}
\includegraphics[width=\textwidth]{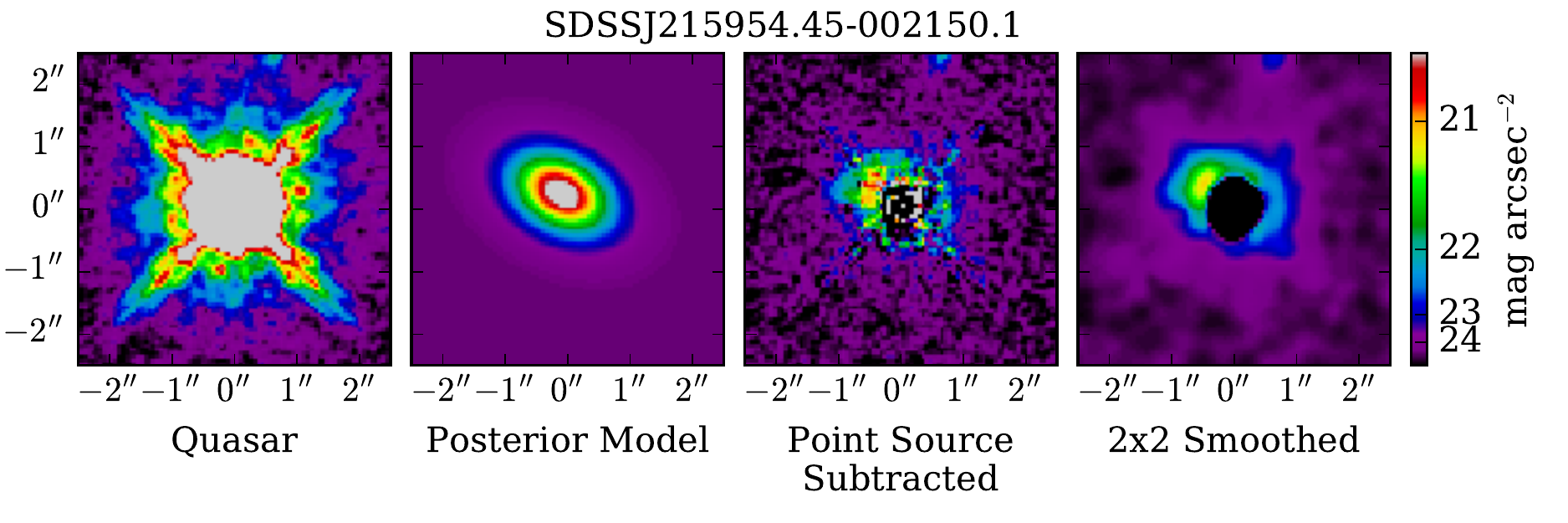}
\includegraphics[width=\textwidth]{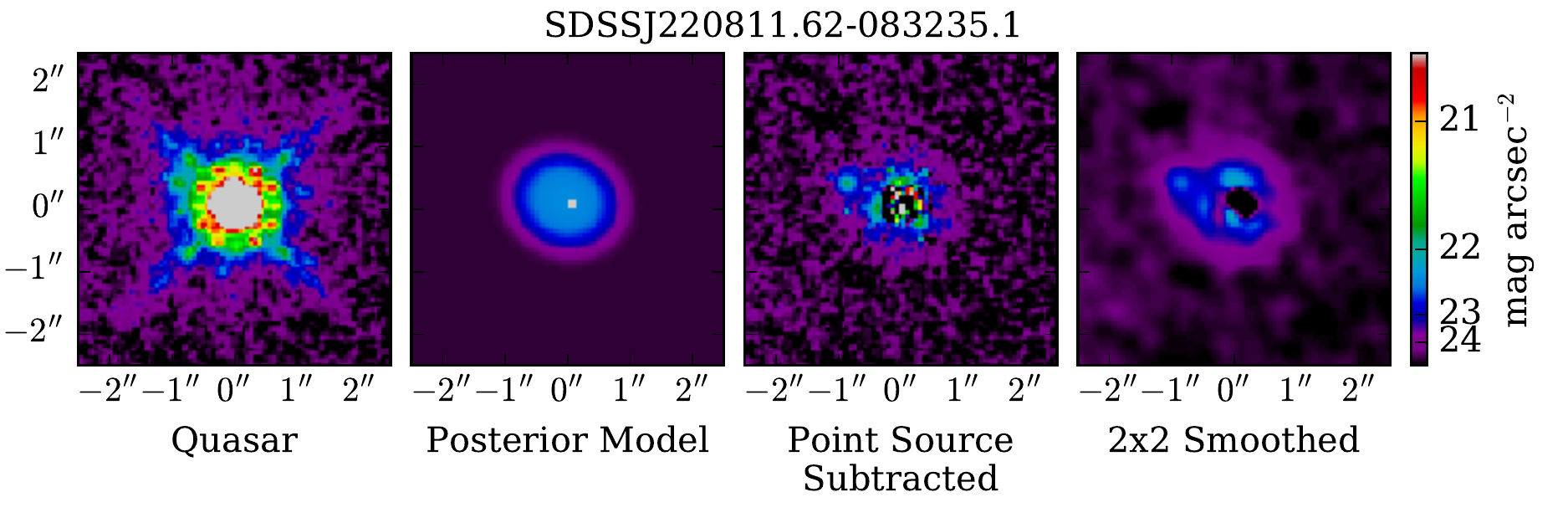}
\includegraphics[width=\textwidth]{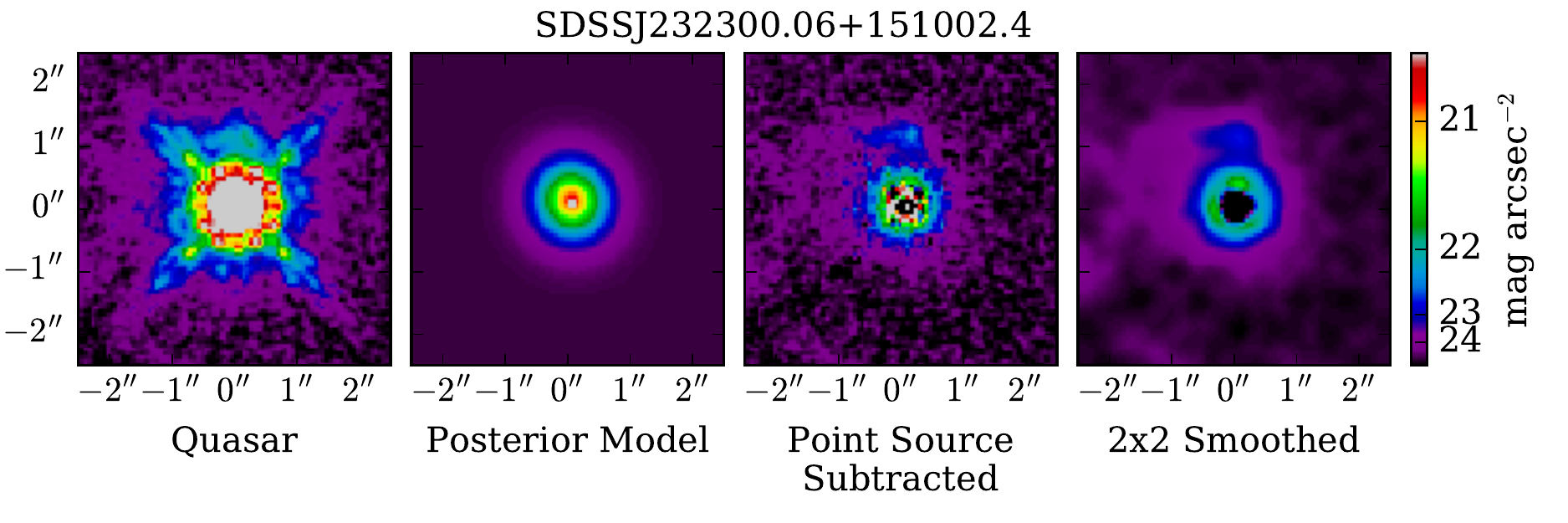}
\caption{(continued) Posterior-weighted models}
\end{figure\flexstar}

\change{Several cosmetic features in these images are worth discussing. In some of the raw model images (``Posterior Model'' panels), a central dot is visible. This is the location of the point source before convolution with the PSF. The point source subtracted images for several quasars (\eg{} SDSS~J113820.35+565652.8 and SDSS~J124949.65+593216.9) show negative cores with a ring-like residual structure at $\simeq0\farcs{}22$ radius (the location of the first maximum in the Airy disk). Others (\eg{} SDSS~J102719.13+584114.3 and SDSS~J155447.85+194502.7) show strong positive core residuals (most apparent in the smoothed images). These features are related to PSF focus and/or SED mismatch rather than significant point source over- or under-subtraction, and also appear when subtracting stars from other stars. The per-pixel S/N is lowest in the center due to the point source shot noise, and these mismatch structures are faint compared to the total point source signal, so they do not significantly affect the fitting and have per-pixel S/N$<1$ in the point source subtracted images.}

Having samples from the full posterior distribution also allows us to examine parameter covariance, as discussed in \sect{bayesian_model}. The strongest of these covariances is between the total magnitude of the point source component and the total magnitude of the S\'ersic component (though we note that the point source magnitude is still determined to within 0.01~mag on average, and the S\'ersic magnitude to 0.1~mag). This is expected since their combined flux must in some sense add up to the observed flux in the WFC3 image, but since some of the total flux is buried in the sky noise the exact form of the covariance function varies from object to object (arguing against removing one of the magnitudes a free parameter from the model). \fig{covariance} plots the posterior distribution for a typical object, marginalizing over each pair of parameters.

\begin{figure\flexstar}
\centering
\includegraphics[width=\textwidth]{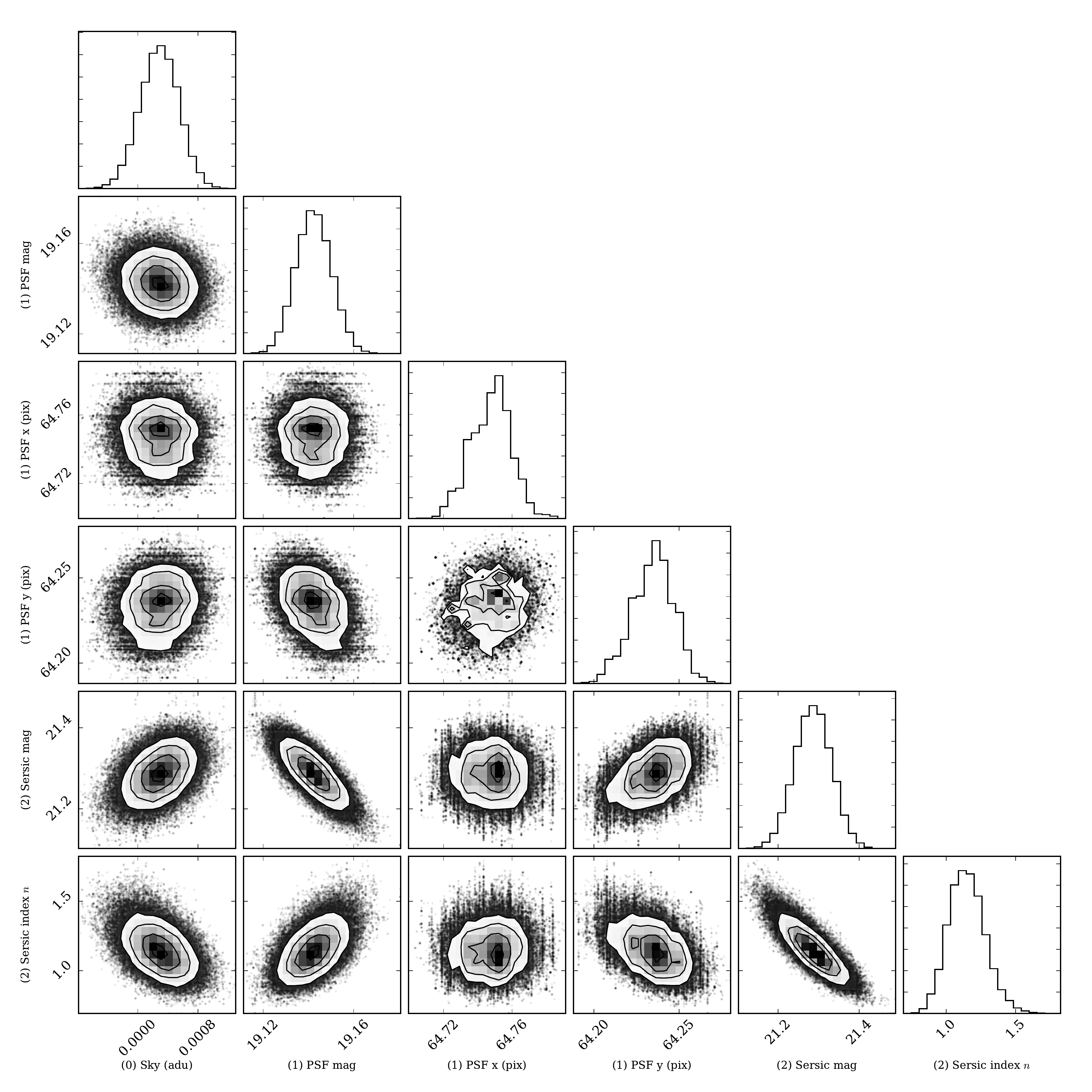}
\caption{Posterior probability distribution for SDSS~J094737.70+110843.3 (200,000 retained samples), showing examples of parameter covariance. Histograms at the top of each column show the 1D marginalized PDF for each free parameter in the model, while contour plots show the 2D marginalized PDF for each pair of parameters. Noise related to image sampling is apparent in the $x,y$ positions of the PSF component, at the 1/100 pixel level. Strong covariance is apparent between several parameters (\eg{} PSF and S\'ersic magnitudes, S\'ersic magnitude and index, S\'ersic major and minor axes). \change{Note: Due to the number of individual panels, this triangular figure is split into 3 sub-figures. These are the upper left panels.}
}
\label{fig:covariance}
\end{figure\flexstar}

\begin{figure\flexstar}
\ContinuedFloat
\centering
\includegraphics[width=\textwidth]{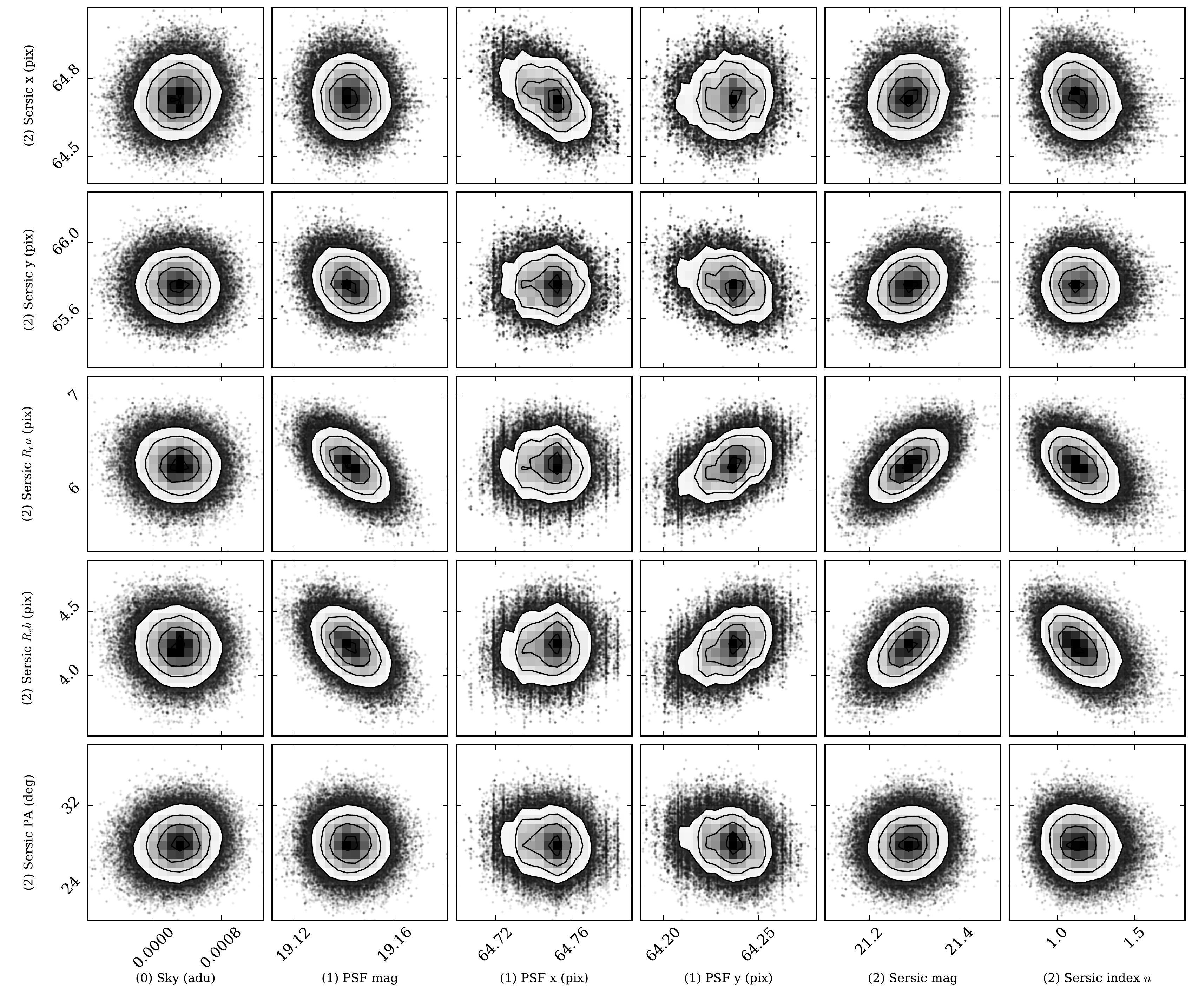}
\caption{(continued) Posterior probability distribution, \change{lower left panels.}}
\end{figure\flexstar}

\begin{figure\flexstar}
\ContinuedFloat
\centering
\includegraphics[width=\textwidth]{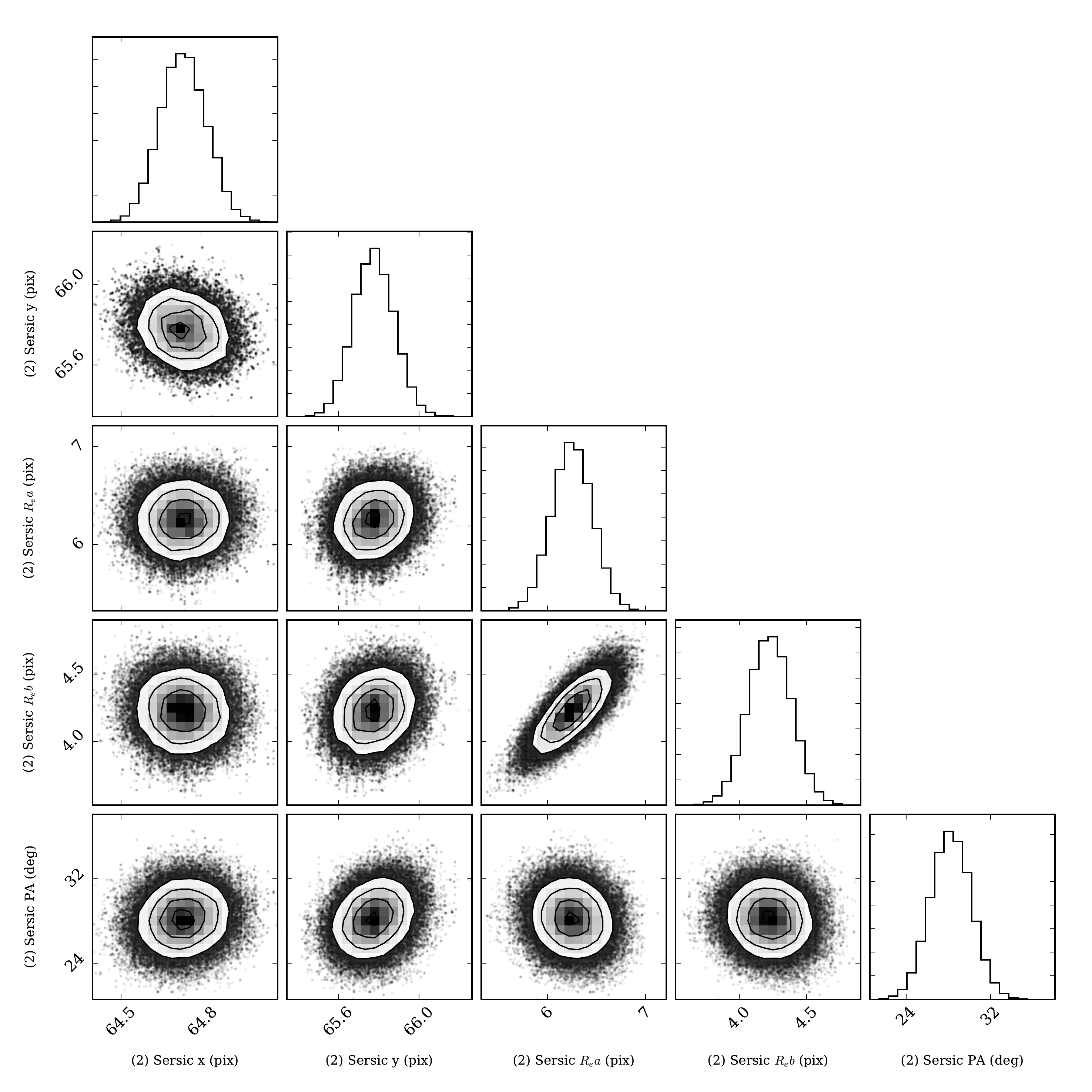}
\caption{(continued) Posterior probability distribution, \change{lower right panels.}}
\end{figure\flexstar}

\subsection{Host Galaxy Photometry}
\label{sec:host_photometry}

For non-morphological analysis (see \sect{stellar_masses} below), we measure revised $F160W$ magnitudes for all objects, rather than using the S\'ersic approximations resulting from the MCMC process. This is primarily important for deriving upper limits for the non-detections, since in those cases the S\'ersic approximation often fits PSF mismatch structures rather than host flux (see \eg{} the posterior model image for SDSS~J135851.73$+$540805.3). \change{Fluxes for the 16 detected galaxies, on the other hand, are identical to their S\'ersic fits, within the 1-sigma magnitude errors.} We perform isophotal photometry on the point source-subtracted images, using a threshold of S/N$= 0.9$ times the RMS background noise to determine the lowest isophote. \change{In the galaxy centers, the high shot noise --- including contributions from the removed quasar point source and the model PSF --- results in few pixels having significant flux values (whether positive or negative, \ie{} S/N is lowest in the center), and are primarily due to PSF mismatch (see discussion in \sect{posterior_analysis} above).} We therefore use values from the model for pixels within a $0\farcs6$ diameter circle surrounding the point source location. For galaxies with total S/N$<2.0$ (the three non-detections), we report the $2\,\sigma$ upper limit. The difference compared to the S\'ersic fit magnitudes is $\Delta{}m<0.5$~mag for the detected objects, but for non-detections the upper limits are $1.0-1.5$~mag brighter than the values indicated by the fit. The observed $F160W$ magnitudes and corresponding $V$-band absolute magnitudes are reported in \tab{quasars_ranked} in Appendix~\ref{sec:appendix_distort}. The host-to-nuclear luminosity ratio $L_\mathrm{Host}/L_\mathrm{Nuc}$ of our sample spans the range from $<0.01$ to $0.16$, with median $0.058$ ($m_\mathrm{Nuc}-m_\mathrm{Host}=-3.2$~mag). This is roughly the range predicted for \change{quasars} with $L/L_\mathrm{Edd}\gtrsim0.1$ from SDSS spectral decompositions \citep[which are sensitive only to host fractions $>0.1$,][]{vanden_berk_spectral_2006}.

\section{Inactive Galaxy Comparison Sample}
\label{sec:comparison_sample}
\subsection{Comparison Sample Selection}
\label{sec:comparison_selection}

As discussed in \sect{intro}, any study that wishes to make a definitive statement about particular AGN triggering mechanisms needs to compare the AGN hosts to a matched sample of inactive galaxies. This matching in principle should be done for all properties except accretion rate (AGN vs.\ non-accreting) and the property we wish to test for (morphology). Our quasars were selected by SMBH mass, but this is unfortunately not a directly measurable quantity for large samples of inactive \zeq{2} galaxies in HST extragalactic survey fields. Stellar masses ($M_{*}$) are, on the other hand, fairly well-constrained in such fields from multi-wavelength SED fitting, when the available data cover the entire rest-frame ultraviolet to rest-frame near-infrared range (with the standard caveats of stellar population synthesis modeling, including an assumed universal initial mass function). The question is then how to select galaxies whose stellar masses match the distribution of stellar masses for the quasar hosts. One approach is to assume a certain $M_\mathrm{BH}-M_{*}$ relation \citep[\eg{}][]{haring_black_2004, kormendy_coevolution_2013}, taking into account scatter in the relation, possible redshift evolution, and biases introduced by our selection function, and then select galaxies in the resulting stellar mass range. The second approach is to match samples using some proxy for $M_{*}$ that is directly observable in both datasets, namely the total $F160W$ magnitude. We opted for this second option, since it requires no assumptions regarding SMBH scaling relations and still captures the full range of galaxies that might reasonably host the high-mass SMBHs. We discuss scaling relations in more detail in \sect{stellar_masses}.

We obtained $F160W$ images for our comparison sample from the CANDELS HST multi-cycle treasury program \citep{grogin_candels:_2011, koekemoer_candels:_2011}, selected using  the redshift and photometry catalogs from the 3D-HST program \citep[v4.1,][]{brammer_3d-hst:_2012, skelton_3d-hst_2014}. The five CANDELS treasury fields represent the best comparison dataset since they image a moderately wide area in $F160W$ (0.22 deg$^2$), and contain a wealth of ancillary data at other wavelengths to ensure accurate redshift determinations, stellar masses, and AGN identifications. We first selected all galaxies in the redshift range \zeq{1.8-2.2}. Because the CANDELS fields also have deep $3.6-8.0\mu{}$m imaging from the \emph{Spitzer Space Telescope}, which samples the red side of the Balmer/4000\AA{} break at \zeq{2}, the 3D-HST photometric redshifts are sufficiently accurate \citep[$\Delta{}z/(1+z) < 0.1$ at \zeq{2},][]{skelton_3d-hst_2014}. We next exclude probable AGN from the sample by cross-matching  the \zsim{2} galaxies with X-ray catalogs \citetext{Subaru/XMM-Newton Deep Survey for UDS, \citealp{ueda_subaru/xmmnewton_2008}; Chandra Source Catalog for the other fields, \citealp{evans_chandra_2010}}. The number of X-ray sources removed from the \zsim{2} sample in each field were: AEGIS: 9, COSMOS: 5, GOODS-N: 4, GOODS-S: 6, UDS: 1. We examined each remaining object and excluded any spurious detections (stars and diffraction spikes). This left us with a parent sample of 1123 \zsim{2} galaxies with $m_{F160W} < 23$~mag, and 150 with $m_{F160W} < 22$~mag.

We then drew 10 samples from the posterior distribution of S\'ersic component magnitude for each of the 16 detected quasar hosts\footnote{Three of the 19 quasars --- SDSS~J082510.09+031801.4, SDSS~J102719.13+584114.3, and SDSS~J135851.73+540805.3 --- had S\'ersic component magnitudes consistent with the noise limit in the images, \ie{} were formally undetected. We did not select comparison galaxies for these objects.}, and selected the \zsim{2} galaxies which most closely matched these samples in $F160W$ magnitude. The quasar hosts are luminous (median $m_{F160W}=21.36$~mag), so there are not enough objects in CANDELS to ensure each host has 10 unique magnitude-matched comparison galaxies. We therefore allowed galaxies to be re-used as comparisons for more than one quasar (\ie{} drawn galaxies were replaced before selecting galaxies matched to the next quasar). In total, 48 comparison galaxies were used once, 18 were used twice, and 18 three or more times (with a maximum of seven times). This gave us 10 redshift- and magnitude-matched inactive comparison galaxy images for each quasar, or a total of 160 images of 84 unique galaxies. \change{We were originally prepared to either down-select to fewer comparison galaxies per quasar or employ weighting in analyses, but decided this was unnecessary since the weighted and un-weighted stellar mass distributions were not significantly different (see \sect{stellar_masses} below). Our analyses simply consider the sample of 84 comparison galaxies without re-use unless otherwise stated (though see Appendix~\ref{sec:appendix_distort} for some consistency checks that were enabled by this re-use).}

\subsection{Addition of Synthetic Quasar Point Sources}
\label{sec:fake_point_sources}

To allow a fair comparison between quasar hosts and inactive galaxies, we needed to ensure that merger signatures were equally detectable in both sets of images. The CANDELS $F160W$ images are deeper than those in our SNAP program, and the quasar images retain systematic residual patterns from slight PSF mismatch, despite our best efforts at PSF matching (see \sect{psf_model}). We therefore added synthetic point sources to the comparison galaxy images and performed the same MCMC point source subtraction procedure used for quasars.

For galaxies that were used as comparisons for multiple quasars (\sect{comparison_selection} above), the images were first transformed with a unique sequence of mirror-flips and 90-degree rotations, so that no two images of a galaxy were exactly the same. For each comparison galaxy image, we then randomly selected one of the eight PSF stars (see \sect{psf_model}) to act as the synthetic point source. This was then scaled to match the magnitude of the subtracted point source for the galaxy's corresponding quasar --- recalling that ten comparison galaxies were selected to match each quasar host --- and inserted into the image centered at the galaxy flux centroid. We then measured the sky noise in the resulting image (the point source having contributed some noise), \change{and added noise if necessary to match the sky noise of the quasar image. This was approximated as an uncorrelated gaussian field rather than correlated poisson noise like the real quasar images, but this added field is not the dominant source of noise in the final images. Of the 160 images, 116 required no added noise, with the other 44 requiring at most $\sigma=0.0047$~e$^-$s$^{-1}$ additional noise, corresponding to 29\% of the final variance after point source subtraction.} Weight images were produced by summing the individual variance sources for these images: the CANDELS (sky) variance map, the CANDELS object signal (approximated as the image in electron units), the scaled point source variance map, and the variance of the added gaussian field. We then performed point source subtraction using the same MCMC parameter estimation procedure described in \sect{model_comps}, supplying only the remaining seven stars as PSF models, \ie{} the star used for each galaxy's synthetic point source was excluded. The resulting galaxy images then contained the same structural residuals from point source subtraction as the quasar host images.

\section{Distortion Ranking and Merger Fraction}
\label{sec:ranking_and_merger_fraction}
\subsection{Expert Ranking Procedure}
\label{sec:expert_ranking}

After preparing the comparison sample, we now had images for a joint sample of 179 quasar hosts and inactive galaxies with similar noise properties and residuals from point source removal. Rather than classifying the galaxies into bins of merger strength or interaction stage \citep[\eg{}][]{cisternas_bulk_2011, kocevski_candels:_2012}, we instead had ten galaxy morphology experts\footnote{\change{Ranking was performed by co-authors Cisternas, Cohen, Hewlett, Jahnke, Mechtley, Schulze, Silverman, Villforth, van der Wel, and Windhorst}} \emph{rank} the 179 galaxies by morphological evidence of distortion due to strong gravitational interactions. Since ranking is a relative measure of interaction strength, it gives a natural way to avoid personal statistical biases (\eg{} what an individual considers a major vs.\ a minor merger), and allows us to examine how our conclusions change as a function of where exactly the final distinction between merger and non-merger is drawn. While ranking the galaxies, these experts were also asked to note any instances where the galaxy was undetected (\ie{} was below the noise limit or completely obscured by image artifacts from the point source subtraction).

We then combined the ranked lists to establish a single consensus sequence of galaxies from the most to the least distorted. \change{This is more difficult than might naively be assumed, since it is equivalent to a ranked voting system and results from social choice theory apply. In particular, Arrow's Impossibility Theorem \citep{arrow_difficulty_1950} states that there is no way to combine such ranked lists such that the consensus sequence has three desirable properties: 1) non-dictatorship, such that each voter receives equal weight, 2) unanimity, such that if all voters agree option $X>Y$, then the consensus sequence also has $X>Y$, and 3) independence of irrelevant alternatives, such that the consensus preference between $X$ and $Y$ depends only upon the individual preferences between $X$ and $Y$, and not relationships between other options. Several techniques are available for relaxing one or more of these properties --- we chose to relax non-dictatorship by allowing de-weighting of individual votes, clipping those that significantly disagreed with the majority for each galaxy.}

The individual lists were merged by calculating the mean rank for each galaxy and its associated variance. We then clipped any individual ranks that were more than $2\,\sigma{}$ from the mean rank for each galaxy. This excluded 55 of the 1790 individual rankings, \change{roughly evenly distributed between rankers (between 0 and 11 individual ranks clipped for each ranker)}. A total of 237 individual rankings were additionally flagged as non-detections. For comparison galaxies that were included multiple times (see \sect{comparison_sample}), we used the inverse variance-weighted mean of the individual rankings as the final rank. If half or more of the individual rankings flagged a galaxy as non-detected, we excluded it from further analysis (as noted in the merger statistics below). Images of all the galaxies in consensus rank order are included in Appendix~\ref{sec:appendix_distort}.

\subsection{Merger Fraction Determination}
\label{sec:merger_fraction}

\change{As with all studies based on visual inspection, individual galaxies show more or less evidence for major mergers. We made this explicit by ranking them on a continuum from most to least evidence. An alternative to merger fractions, then, is to consider the two samples' distributions within this ranked continuum. These distributions are formally indistinguishable under either a 2-sample Kolmogorov-Smirnov test ($D=0.21, p=0.45$) or 2-sample Anderson-Darling test ($A^2=-0.38, p=0.52$). The notion of categorizing galaxies as mergers or non-mergers is convenient for some of the discussion in \sect{discussion}, so we discuss the derivation of a merger fraction for each sample below.
}

To \change{estimate} the quasar host and inactive galaxy merger fractions we visually inspected the consensus sequence and selected a particular cutoff rank, below which we could no longer find any clear merger signatures. We selected rank \cutoffrank{} for this cutoff (corresponding to 32\% of the final sample of all quasars and inactive galaxies, minus the non-detections), but note that the particular choice of rank may differ among individual experts. In Appendix~\ref{sec:appendix_distort} we explore in detail how this choice affects the inferred merger fractions. We only note here that for any reasonable choice of cutoff rank, the qualitative aspects and conclusions below do not change.

\begin{figure}
\centering
\includegraphics[width=\columnwidth]{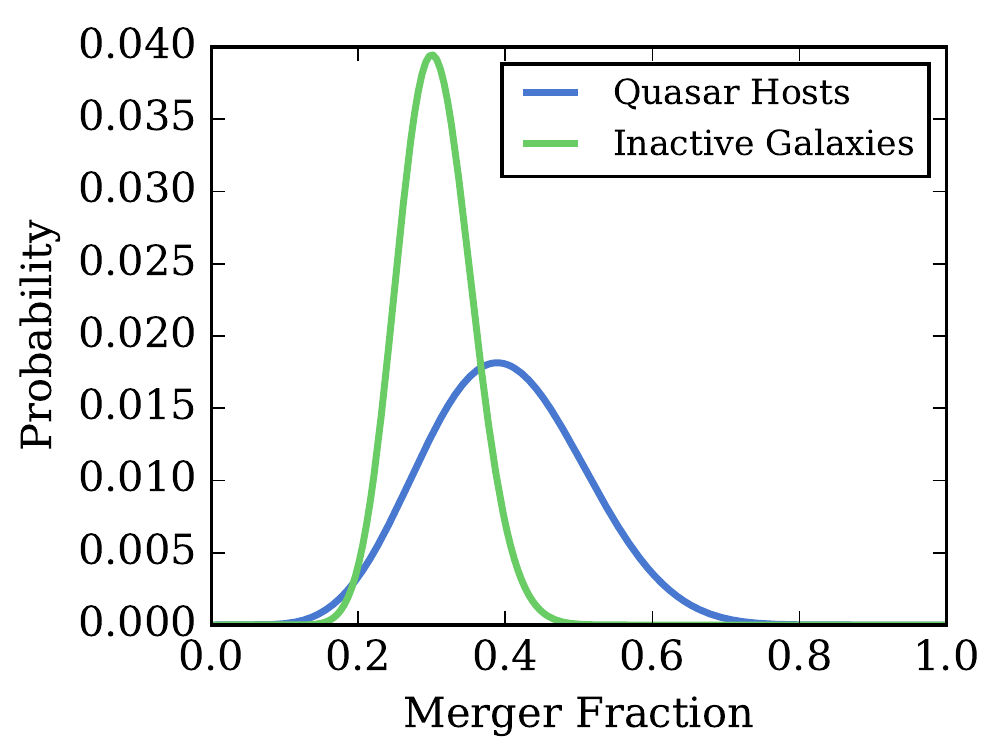}
\caption{Probability distributions for the inferred merger fractions of both quasar hosts and inactive galaxies. The inactive galaxy distribution peaks at \fracgal{}, and the quasar distribution peaks at \fracqso{}. Neither has been corrected for observational biases, which by design are the same for both samples, so we caution that they should only be interpreted relative to one another, not as intrinsic or absolute merger fractions. The combined uncertainty means that the probability of the quasar fraction being larger than the inactive galaxy fraction is $P(f_\mathrm{m,qso}>f_\mathrm{m,gal}) = \signifprob{}$, \ie{} the slight enhancement is not statistically significant.}
\label{fig:beta_dists}
\end{figure}

With a merger/non-merger cutoff rank of \cutoffrank{}, we identified \qsomergers{} quasar hosts with evidence for major mergers, \qsononmergers{} with no evidence for major mergers, and \qsonondetects{} indeterminate (more than half of individuals flagged them as non-detections). For the inactive galaxies, \galmergers{} had evidence for major mergers, \galnonmergers{} had no evidence for major mergers, and \galnondetects{} were indeterminate. The probability distribution describing the inferred merger fraction given the finite number of experimental trials is the beta distribution. The probability density function (PDF) for the beta distribution is given by \eqn{beta_dist}:
\begin{equation}
P(x;a,b) = \frac{(a+b+1)!}{a!\,b!}x^{a}(1-x)^{b}
\label{eqn:beta_dist}
\end{equation}
Here $a$ and $b$ are integers denoting the number of mergers and non-mergers, respectively. The resulting probability distributions for both quasar hosts and inactive galaxies are shown in \fig{beta_dists}. The modes or peaks of the distributions are simply the usual merger fractions, $a/(a+b)$. The inferred parent population merger fractions, with 68\% confidence intervals, are thus $f_\mathrm{m,qso}=\fracqso{}\pm{}0.11$ for quasar hosts, and $f_\mathrm{m,gal}=\fracgal{}\pm{}0.05$ for inactive galaxies.

\section{Discussion}
\label{sec:discussion}
\subsection{Comparison of Inferred Merger Fractions}
\label{sec:compare_merger_fracs}

Although we use the term ``merger fraction,'' we emphasize that the distorted fractions above \emph{should not} be used as absolute merger fractions to compare to other studies of high-mass galaxies. We, as with other studies, have invariably missed a few real mergers and possibly mis-identified some non-mergers due to noise and residual patterns associated with the point source subtractions (see discussion in Appendix~\ref{sec:appendix_images}). Rather, they should only be interpreted \emph{relative} to each other since both samples have the same observational limitations. The probability distribution for the quasar merger fraction in \fig{beta_dists} rules out the extreme scenario immediately. The merger fraction is not consistent with values near 1 (99.7\% confidence interval: $0.13-0.72$), as might be expected if every massive quasar were growing due to ongoing merger activity.

The quasar host PDF peaks at a slightly higher merger fraction than the inactive galaxy PDF, corresponding to a merger fraction enhancement of $f_\mathrm{m,qso} / f_\mathrm{m,gal} = 1.3$. The enhancement is not significant, however. Given two random variables $X$ and $Y$, the probability that $X>Y$ is the integral of the joint PDF over the region where this inequality holds. For the two distributions in \fig{beta_dists}, this probability is $P(f_\mathrm{m,qso}>f_\mathrm{m,gal}) = \signifprob{}$, or $\signifsigma{}\,\sigma$. We can invert this to examine the statistical sensitivity of our observations --- given the fixed quasar sample size, and assuming the observed PDF for the inactive galaxies, we can calculate the minimum merger fraction that would have resulted in a significant ($>3\,\sigma$) signal. For 18 quasars, this would have required 13 mergers, for a merger fraction of $f_\mathrm{m,qso} \geq 0.72$ or enhancement of $f_\mathrm{m,qso} / f_\mathrm{m,gal} \geq 2.4$. Correspondingly, if the true intrinsic distortion fractions are \fracgal{} and \fracqso{}, we can ask how much larger the sample would need to be to detect this enhancement signal at a significant level. \emph{Both} the quasar and comparison galaxy sample sizes would need to be increased by a factor of $7.6$ to detect an enhancement signal at $2\,\sigma{}$, and a factor of $17.2$ for $3\,\sigma{}$. This underscores the statistical insignificance of the observed enhancement, and the need for cautious interpretation when dealing with beta-distributed quantities inferred from small samples.

\subsection{Are These High-Mass Quasars Preferentially Hosted in Mergers?}
\label{sec:merger_preference}
Although the observed enhancement is not significant, the data are still consistent with the quasars having either a slightly enhanced merger fraction, or no enhancement. The data are not, however, consistent with the quasars having a large merger fraction enhancement. If the observed enhancement is real (\ie{} if it were still present with much larger samples), it could indicate that mergers are simply one of several possible AGN triggering mechanisms, rather than the dominant or only mechanism. 

One possible caveat is that of significant time lags between the appearance of merger signatures and the onset of quasar activity. That is, if gas disturbed or provided by a merging companion takes long enough to reach the SMBH (\ie{} several times the dynamical timescale), or is first reprocessed via an episode of star formation, morphological signatures of a merger may no longer be observable. Observational evidence for such time lags is necessarily indirect, and the most successful studies have examined growth regimes very different from those in our study --- \eg{} \citet{wild_timing_2010}, who studied AGN 2.5 orders of magnitude less massive than ours at \zeq{0.01-0.07}, \change{and inferred a time lag of 250~Myr between peak starburst activity and peak AGN accretion}. Theoretical models of merger-induced AGN activity predict a delay of $\simeq100$~Myr between galaxy coalescence and the peak of quasar activity \citep{di_matteo_energy_2005, hopkins_cosmological_2008}. These estimates are at any rate shorter than the timescale over which morphological merger signatures are still visible \change{\citep[as high as $\gtrsim{}1$~Gyr for gas-rich mergers][]{lotz_effect_2010, lotz_effect_2010-1}}. For exceptionally long time lags, such that merger signatures are almost completely erased, the observational signatures would become essentially indistinguishable from violent disk instabilities \citep[VDI, \eg{}][]{bournaud_observed_2012, trump_no_2014}. In such a case the theoretical model might even be indistinguishable. That is, if gas transport times are several times longer than the dynamical time, is the merger responsible for the loss of angular momentum, or are secular processes like VDI simply acting upon the gas-rich merged galaxy? For the specific feeding mechanism we are testing --- near-zero angular momentum gas provided directly by a major merger --- the relevant timescale should be closer to the free-fall time and thus shorter than the timescale for merger signatures to be smoothed out.

\begin{figure}
\centering
\includegraphics[width=\columnwidth]{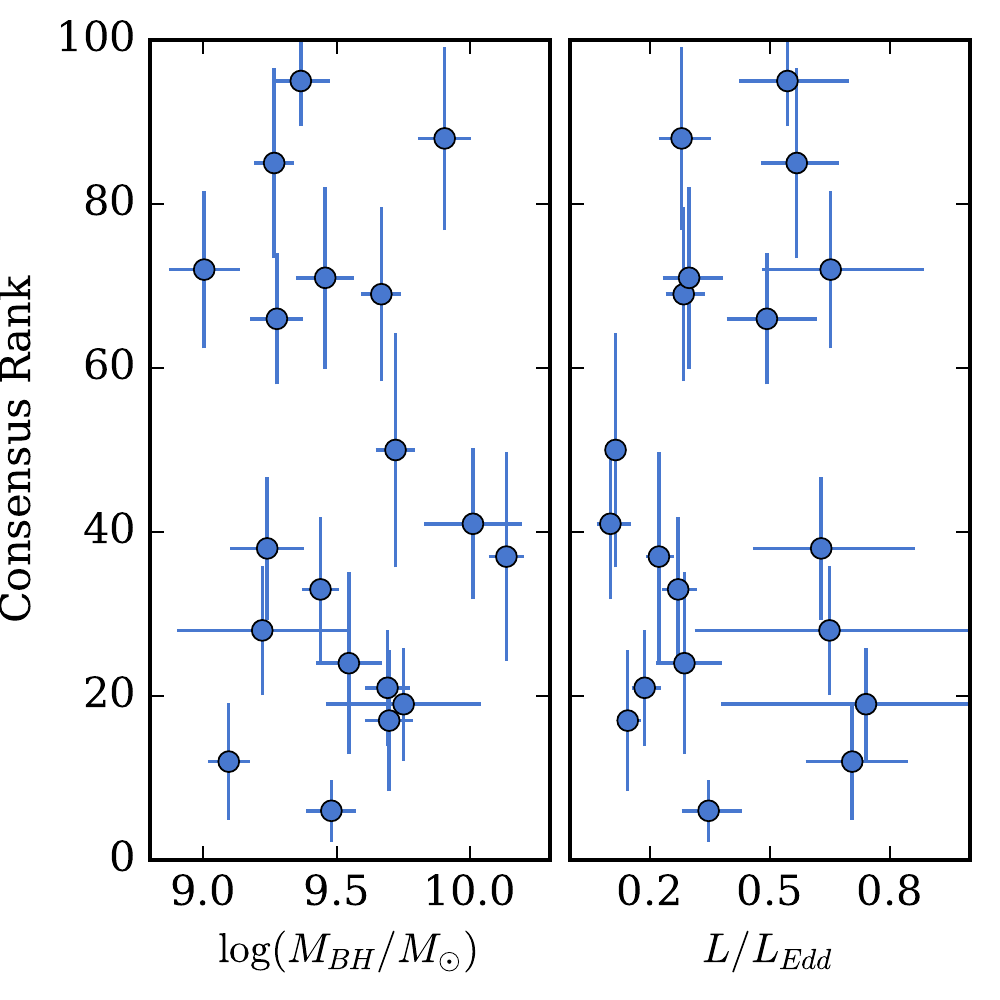}
\caption{\change{Host distortion rank as a function of black hole properties. Left: Black hole mass. Right: Eddington ratio $L/L_\mathrm{Edd}$. Vertical error bars are the standard error on the mean rank. Horizontal error bars are from the stated uncertainty on $M_\mathrm{BH}$ from \citet{shen_catalog_2011}. No trend in distortion rank is observed as a function of either black hole property.}
}
\label{fig:accretion_rate}
\end{figure}

\change{We also examined the incidence of merger features as a function of black hole mass and accretion rate (\fig{accretion_rate}), with the caveat that our sample spans only about a factor of ten in each. No trend is observed in distortion rank as a function of black hole mass, and thus no evidence that the higher-mass black holes are nearer the end of a current merger. Similarly, no trend is observed as a function of either luminosity or Eddington ratio, as might be expected if nuclear gas availability were systematically greater for merger-triggered quasars. We have planned a future study to address the question of merger triggering in high-Eddington systems specifically, as contrasted with low-Eddington AGN and inactive galaxies.
}

\subsection{Results from \change{AGN} Environment Studies}
\label{sec:comp_environment}
Environmental diagnostics provide another test of merger or interaction triggering hypotheses. Small-scale environmental studies test a slightly different hypothesis compared to our study --- essentially, that interactions trigger gas instabilities within a galaxy at early merger stages or in non-merging close encounters. Studies of small-scale clustering of quasars \citep{hennawi_binary_2006, myers_quasar_2008} found some evidence for an excess in the quasar spatial autocorrelation function at small ($\simeq10$~kpc) scales. They attributed this excess to gravitational interaction events triggering quasars in rich environments. \citet{silverman_environments_2009} examined the AGN fraction in galaxies as a function of their local environmental density using a nearest-neighbors approach. They found that the hosts \change{of these lower-luminosity AGN} generally trace the same environments as star-forming galaxies (both processes requiring gas), with a preference toward under-dense regions \change{for AGN hosts comparable in mass to our inactive sample ($M_{*}\gtrsim{}10^{11}$~\MSun{})}.

A related diagnostic for examining AGN triggering via gravitational interactions is the study of galaxies in kinematic pairs, \ie{} those that are close in mass, spatial projection, and line-of-sight velocity, looking once again for an enhancement to the number of AGN in close pairs versus a field sample. \citet{ellison_galaxy_2008} studied low-redshift ($z<0.15$) kinematic pairs from SDSS (2402 galaxies in pairs, 69583 in the field control sample), finding the \change{(lower-luminosity)} AGN fraction in pairs to be consistent with that of the control sample. \citet{silverman_impact_2011} performed a similar study at \zeq{0.25-1.05} using the zCOSMOS spectroscopic sample (562 galaxies in pairs, 2726 control galaxies). They detected significant differences in the AGN fractions of three subsamples: close pairs (9.7\%), wide pairs (6.7\%), and isolated galaxies (3.8\%). They estimated that 17.8\% of moderate-luminosity nuclear activity is triggered during early-stage interactions, leaving the further $\simeq80$\% unaccounted for. Both the SDSS and zCOSMOS studies posited that later stages of major mergers may account for some of the missing triggers. \citet{lackner_late-stage_2014} examined this by looking for galaxies with multiple nuclei at small separations (\ie{} pairs no longer distinguishable as such in the low-resolution ground-based imaging), finding that the combination of wide pairs, close pairs, and late-stage mergers account for a total of 20\% of AGN activity at \zeq{0.25-1.0}. This is then consistent with the \citet{cisternas_bulk_2011} study, which found that morphologically-identified major mergers are not a dominant trigger for \change{low-to-moderate luminosity AGN}. We have now further shown that such major mergers are not a dominant triggering mechanism for the high-mass quasars that dominate SMBH accretion at \zeq{2}.

\subsection{Results from Red and Dust-Obscured Quasar Studies}
\label{sec:comp_red}
\change{As mentioned in \sect{intro}, studies of reddened and dust-obscured quasars have generally found very high merger fractions, in contrast with most studies of unobscured AGN or quasar populations mentioned above. Various obscured or dust-rich quasar selection methods target slightly different spectral features. The first HST studies of red quasars targeted objects with far-infrared excesses \citep[\eg{}][]{canalizo_quasistellar_2001}, finding evidence for major mergers in 8 of 9 hosts. Although the authors argue that chance hosting is unlikely --- \ie{} the quasar and starburst activity are related --- such objects make up a small fraction of low-redshift quasars, so correspondingly represent a small contribution to the total triggering budget.

\citet{zakamska_type_2006} selected Type~2 quasars at \zsim{0.2-0.4} from SDSS using the scattered emission-line luminosities as a proxy for total nuclear luminosity --- \ie{} objects expected to be similar to moderate-luminosity Type~1 quasars, but viewed from an angle where the circumnuclear dust torus obscures the direct AGN light, so not distinct from an evolution standpoint. They found evidence for a major merger in only 1 of 9 hosts, and evidence for tidal debris in 3 of 9, roughly in line with the lower merger fractions found in low-redshift Type~1 quasars \citep{mclure_comparative_1999, dunlop_host_2001, dunlop_quasars_2003, floyd_host_2004}.

More recently, considerable effort has gone into the study of dust-reddened Type~1 quasars  selected from a combination of radio and near-IR data \citep[F2M quasars,][]{glikman_first-2mass_2007}. These highly-reddened quasars make up $\simeq{}10$\% of the luminous quasar population, are among the most intrinsically luminous at any given redshift \citep{glikman_first-2mass_2012}, and their host galaxies show a high incidence of mergers \citep[merger fractions of $\simeq{}0.8-0.85$,][]{urrutia_evidence_2008, glikman_major_2015}.

How do we reconcile these conflicting results with our current study and other unobscured quasar host studies? Reddening and obscuration can (at least) originate from non-evolutionary nuclear geometric effects (torus obscuration), evolutionary effects (a buried/blowout phase), and non-evolutionary host geometric effects (\ie{} quasars that happen to be hosted in asymmetric, dust-rich, ULIRG-like galaxies). Any given red selection method may pick up some combination of the three. Type~2 selection like that of \citet{zakamska_type_2006} purports to select only for torus effects, and the similarity of those hosts to Type~1 hosts seems to support this. For red samples with high host merger fractions, additional evidence is needed to determine whether the mergers are an evolutionary feature or a host sub-population feature. The F2M quasars are the best bet for quasars triggered by early-stage galaxy mergers \citep[see especially discussions in][]{glikman_first-2mass_2012, urrutia_spitzer_2012}, but the universality of such triggering remains unclear. We argue in \sect{merger_preference} above that, assuming the usual quasar lifetime estimates, we would see far more evidence for mergers in our sample if merger triggering were universal and all (or even most) quasar hosts begin as trainwreck mergers like the F2M hosts.
}

\subsection{Bonus: Stellar Masses and the $M_\mathrm{BH} - M_{*}$ Relation}
\label{sec:stellar_masses}
\begin{figure}
\centering
\includegraphics[width=\columnwidth]{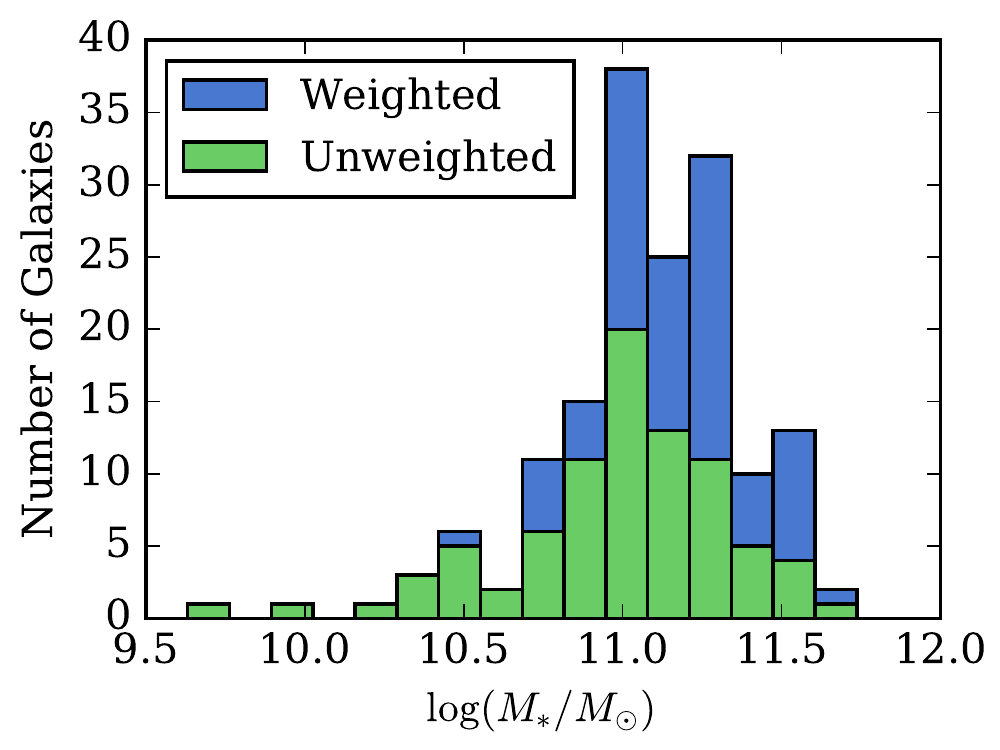}
\caption{Distribution of comparison galaxy stellar masses, from the 3D-HST SED fits. The green, lower histogram shows the intrinsic mass distribution of the comparison sample of 84 galaxies, with median stellar mass $\log(M_{*}/\MSun{})=11.0$. The blue, upper histogram shows the same distribution, where each galaxy has been weighted by the number of times it was used as a luminosity-matched comparison galaxy to a quasar. This weighted distribution has a median mass $\log(M_{*}/\MSun{})=11.1$.}
\label{fig:comp_stellar_masses}
\end{figure}

The distribution of the inactive galaxy stellar masses taken from the 3D-HST catalogs is plotted in \fig{comp_stellar_masses}. Two histograms are shown: one where each galaxy is counted only once, and one where each is weighted by the number of times it was used for comparison to a quasar host. The median stellar mass of the unweighted comparison sample is $\log(M_{*}/\MSun{})=11.0$. The median stellar mass of the weighted sample (\ie{} with the distribution of $F160W$ magnitudes matched to the quasar hosts) is $\log(M_{*}/\MSun{})=11.1$. We note that CANDELS/3D-HST are complete to much lower masses. There is no significant trend in stellar mass as a function of distortion rank, consistent with previous studies finding no significant mass-dependence of the major merger rate over similar stellar mass ranges \citep[\eg{}][]{xu_major-merger_2012}.

We can obtain crude estimates of the quasar host galaxy stellar masses from their $F160W$ magnitudes by adopting some assumptions about their stellar populations. Since we sample the SED at a longer wavelength than the Balmer/4000\AA{} break, the light comes mostly from older stars that account for the bulk of stellar mass. The general procedure is to calculate a galaxy's luminosity from some bandpass, then multiply by a stellar mass-to-light ratio, $M/L$. This assumes a particular SED, which may differ from object to object. Alternatively, we could use the empirical relation (with scatter) between $F160W$ magnitude and stellar mass from the \citet{skelton_3d-hst_2014} catalog, which implicitly encodes the full range of observed SEDs or $M/L$ ratios for galaxies at a given redshift and stellar mass. The quasar hosts and inactive galaxies show a similar range of morphologies, so we take this approach.

Since this is the same population of galaxies from which the comparison sample was drawn, and since the quasar and inactive galaxy samples have the same luminosity distribution by construction, this should result in a mass distribution roughly the same as the inactive galaxies. Indeed, the quasar hosts have a stellar mass distribution of $\log(M_{*}/\MSun{})=11.2\pm0.4$, but with an uncertainty on the mean value of 0.4~dex, due to the intrinsic scatter in the $m_{F160W}-M_{*}$ relation (in turn due to the physical range in SEDs for a fixed magnitude). \change{There is no statistically significant difference between the distributions of stellar mass for the mergers and non-mergers, as diagnosed by either a 2-sample Kolmogorov-Smirnov test ($D=0.33, p=0.68$) or 2-sample Anderson-Darling test ($A^2=0.046, p=0.33$).} The average $M/L\simeq0.5$ is consistent with relatively young stellar populations found in previous quasar host studies \citep[\eg{}][]{jahnke_quasar_2004, jahnke_ultraviolet_2004, sanchez_colors_2004}.

\begin{figure}
\centering
\includegraphics[width=\columnwidth]{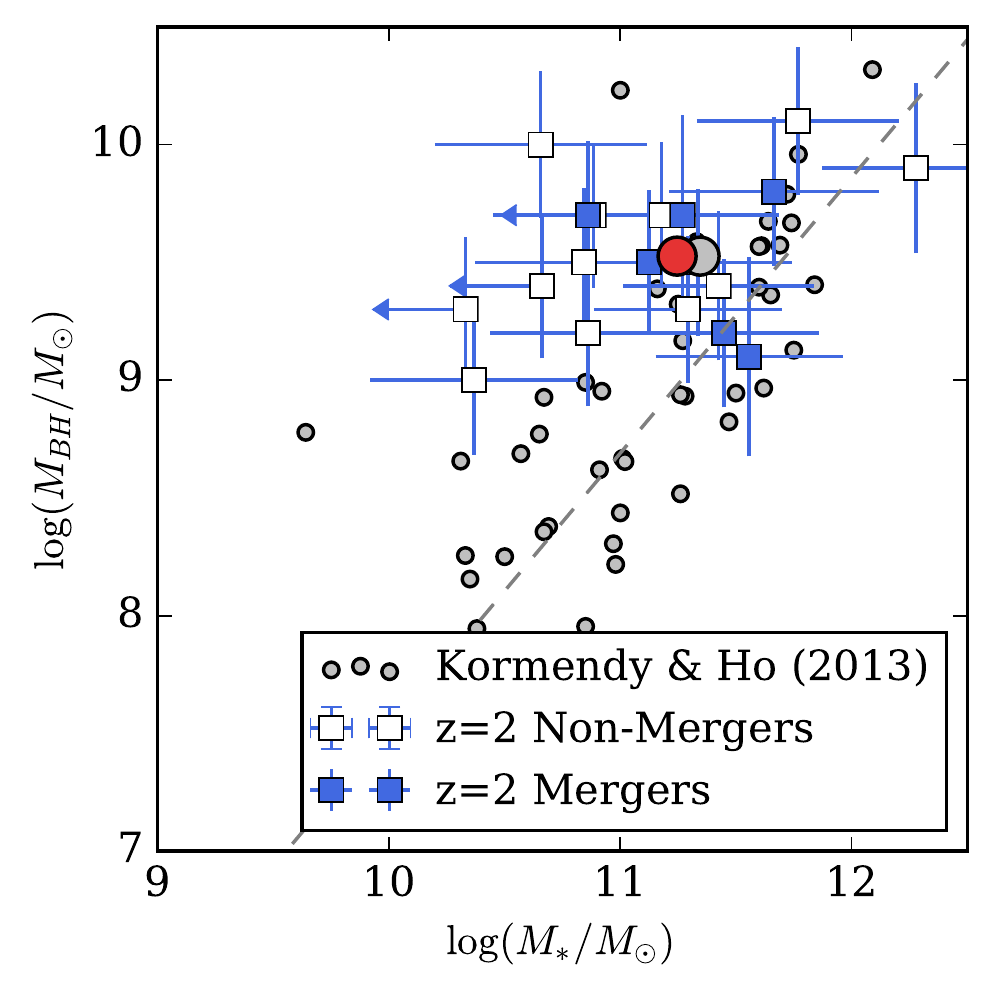}
\caption{Galaxy stellar mass-black hole mass relation. Gray circles are local ellipticals and classical bulges collected from the literature by \citet{kormendy_coevolution_2013}. The dashed gray line is their fit to the \zeq{0} data. Blue squares are the \zeq{2} quasar hosts (filled: mergers, open: non-mergers). Error bars \change{are dominated by} scatter in the virial mass calibrations (for BH masses), and scatter in the \zeq{2} stellar mass-$F160W$ magnitude relation (for stellar masses). The large red circle is the sample mean for the quasar hosts. The large gray circle shows the expected population mean given the biases of our selection function (see text). The observed distribution is thus consistent with the \zeq{0} relation propagated through our selection function.
}
\label{fig:mbh_mstars}
\end{figure}

With stellar mass estimates for the quasar hosts, we can now compare them to the local $M_\mathrm{BH}-M_{*}$ relation. \fig{mbh_mstars} shows the \zeq{2} quasar hosts alongside the local relation derived by \citet{kormendy_coevolution_2013}. The error bars include contributions from the scatter in both the virial black hole mass calibration (0.3~dex) and the $F160W$ magnitude-stellar mass relation (0.4~dex, encoding the range of SEDs for a galaxy with a given magnitude). The median black hole to stellar mass ratio is $\Gamma=\log(M_\mathrm{BH}/M_{*})=-1.7$. However, some bias in this observed ratio compared to the intrinsic relation is expected, given the sample selection function \citep{lauer_selection_2007, schulze_selection_2011}. In particular, when selecting at the high BH mass end of the relation, the corresponding stellar masses are in the exponentially declining regime of the stellar mass function. This leads to average stellar masses lower than the relation. We estimate the expected bias in $\Gamma$ following the framework of \citet{schulze_selection_2011}, using the galaxy stellar mass function from \citet{ilbert_mass_2013}, and the \zeq{2} black hole mass function and Eddington ratio distribution function from \citet{schulze_cosmic_2015}. Assuming the SDSS flux limit, and a BH mass selection limit $M_\mathrm{BH}>10^9$, this predicts a bias of $\Delta{}\Gamma{}=0.37$ over the \citet{kormendy_coevolution_2013} value of $\Gamma=-2.19$ for black holes with $M_\mathrm{BH}=10^{9.5}$~\MSun{}. Our quasar hosts are thus consistent with the local relation ($\Delta\Gamma=0.1$) within uncertainties once selection bias is accounted for, as is the case with previous high-redshift studies \citep{schulze_accounting_2014}.

We note in passing that the lack of redshift evolution is consistent with a picture where the scaling relations arise through non-causal means, \ie{} galaxies approach the cosmic mean $\Gamma{}$ via mergers rather than direct coupling via AGN feedback \citep{peng_how_2007, jahnke_non-causal_2011}. However, a lack of evolution does not itself rule out strong AGN feedback, since feedback models can be constructed that predict weak or no evolution.

\section{Conclusions}
\label{sec:conclusions}
We have performed a study of 19 quasar host galaxies and 84 inactive galaxies at \zeq{2}, having 10 experts blindly rank them by evidence for distortions due to major mergers. The inactive galaxies were luminosity-matched to the quasar hosts and subjected to the same MCMC modeling and point source subtraction procedure, producing comparison images with the same systematic observational biases and limitations. After synthesizing the expert rankings into a consensus distortion sequence, we have demonstrated that the \change{quasar hosts are consistent with being uniformly distributed within the merger sequence of inactive galaxies. The inferred} major merger fraction in host galaxies of massive quasars at \zeq{2} is not significantly higher than the major merger fraction for inactive galaxies, thus the bulk of black hole accretion at the peak of quasar activity is not merger-triggered. This is in line with previous findings regarding the bulk of black hole growth at \zeq{0.3-1.0} \citep{cisternas_bulk_2011} and \change{lower-luminosity AGN} at \zeq{2} \citep{schawinski_hst_2011, kocevski_candels:_2012}, supporting the interpretation that mergers are not the dominant fueling channel by which cosmic black hole mass is built up. We also found no trend in specific accretion rate \change{or black hole mass} as a function of merger rank, \change{over a modest (factor of 10) range in each}.

We also show that, for a reasonable set of assumptions about the stellar populations of the quasar hosts, they have stellar masses that are consistent with the local black hole-to-stellar mass scaling relation once selection biases have been accounted for. This is consistent with \citet{schulze_accounting_2014} who find that previous high-redshift studies also have not observed a significant offset from the local scaling relations, once observational biases have been accounted for.

\acknowledgements{}
\change{We thank the anonymous referee for their careful review and detailed suggestions which improved the manuscript.} MM and KJ acknowledge support through the Extraterrestrische Verbundforschung program of the German Space Agency, DLR, grant number 50~OR~1203. MM acknowledges support provided by NASA through grant GO-12613.011-A from the Space Telescope Science Institute, which is operated by the Association of Universities for Research in Astronomy, Inc., under NASA contract NAS 5-26555. AS acknowledges support by JSPS KAKENHI Grant Number 26800098. \change{This research made use of Astropy, a community-developed core Python package for Astronomy \citep{the_astropy_collaboration_astropy:_2013}.} MCMC corner plots make use of \texttt{corner.py} \citep{foreman-mackey_triangle.py_2014}.

\textit{Facility:} \facility{HST (WFC3 IR)}

\appendix{}
\section{Consensus Distortion Sequence and Choice of the Merger/Non-Merger Cutoff}
\label{sec:appendix_distort}
As discussed in \sect{merger_fraction}, the choice of where to draw the merger/non-merger distinction is somewhat arbitrary and may differ even among galaxy morphology experts. We chose rank \cutoffrank{} as a reasonable fiducial cutoff rank for our discussion, but also examined how the inferred merger fraction distributions were affected by selecting a different cutoff rank. \fig{cutoff_test} shows the result of this experiment for cutoff ranks between 15 and 45. The qualitative interpretation --- that $f_\mathrm{m,qso}$ is slightly higher than $f_\mathrm{m,gal}$, but not significantly --- is essentially independent of cutoff rank for any reasonable choice. That is, for no cutoff rank in the range $15-45$ is the enhancement of $f_\mathrm{m,qso}$ over $f_\mathrm{m,gal}$ significant ($>3\,\sigma$). In \fig{consensus_sequence}, we reproduce the point source-subtracted images of all quasar hosts and inactive galaxies in consensus sequence order, so that readers may make their own assessment of the appropriate cutoff rank. The images are presented with the asinh stretch and color map that was provided to the experts when ordering the sequence. For objects that were ranked multiple times (see \sect{comparison_selection}), the image with the smallest rank variance (\ie{} the image that dominated the consensus rank determination) is shown.

\begin{figure\flexstar}
\centering
\includegraphics[width=\textwidth]{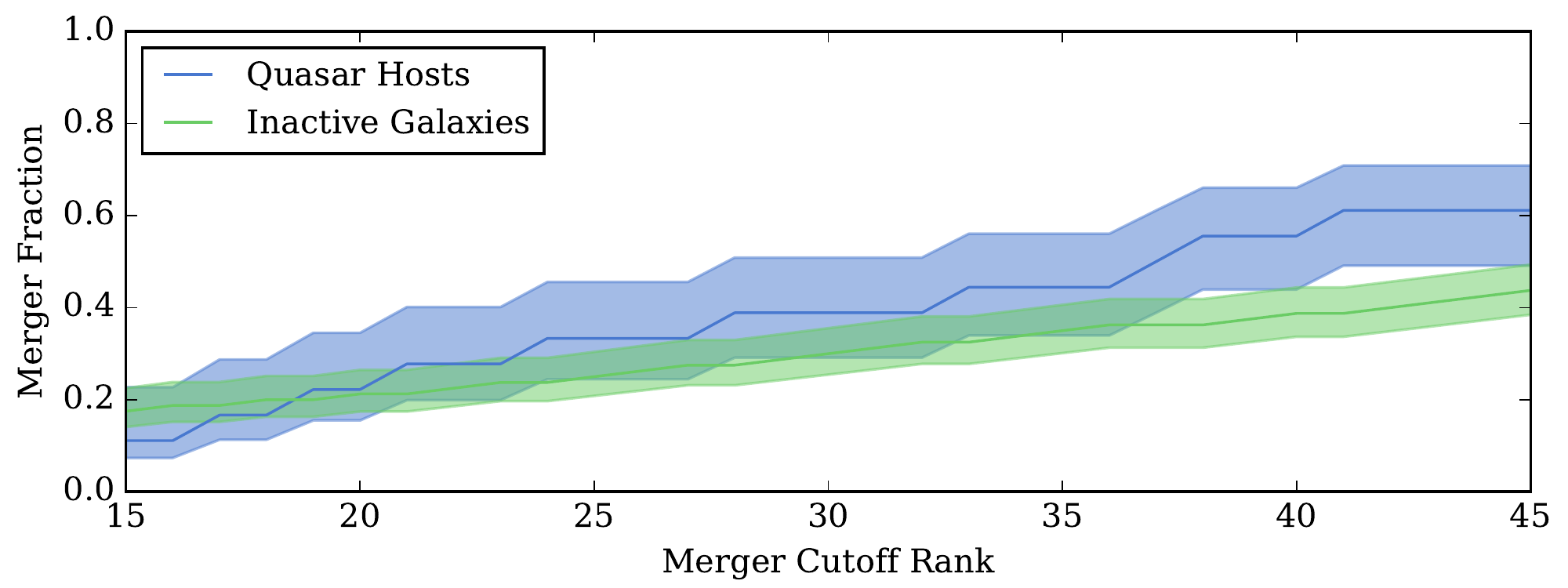}
\caption{Effect of selecting a different merger/non-merger cutoff rank. We selected rank \cutoffrank{} as the fiducial cutoff for our discussion, but note that individual interpretations of merger signatures might select a range of cutoffs near this rank. The red and blue lines show the inferred merger fractions for quasar hosts and inactive galaxies, respectively, as a function of cutoff rank. The shaded regions show the 68\% confidence intervals of the associated beta PDFs. In no case (including the range from $38-45$) is the enhancement of $f_\mathrm{m,qso}$ over $f_\mathrm{m,gal}$ significant ($3\,\sigma$, see \sect{compare_merger_fracs}). The qualitative interpretation --- that $f_\mathrm{m,qso}$ is slightly higher than $f_\mathrm{m,gal}$, but not significantly --- is thus independent of cutoff rank, for any reasonable cutoff rank.
}
\label{fig:cutoff_test} 
\end{figure\flexstar}

\begin{figure}
\centering
\includegraphics[width=\columnwidth]{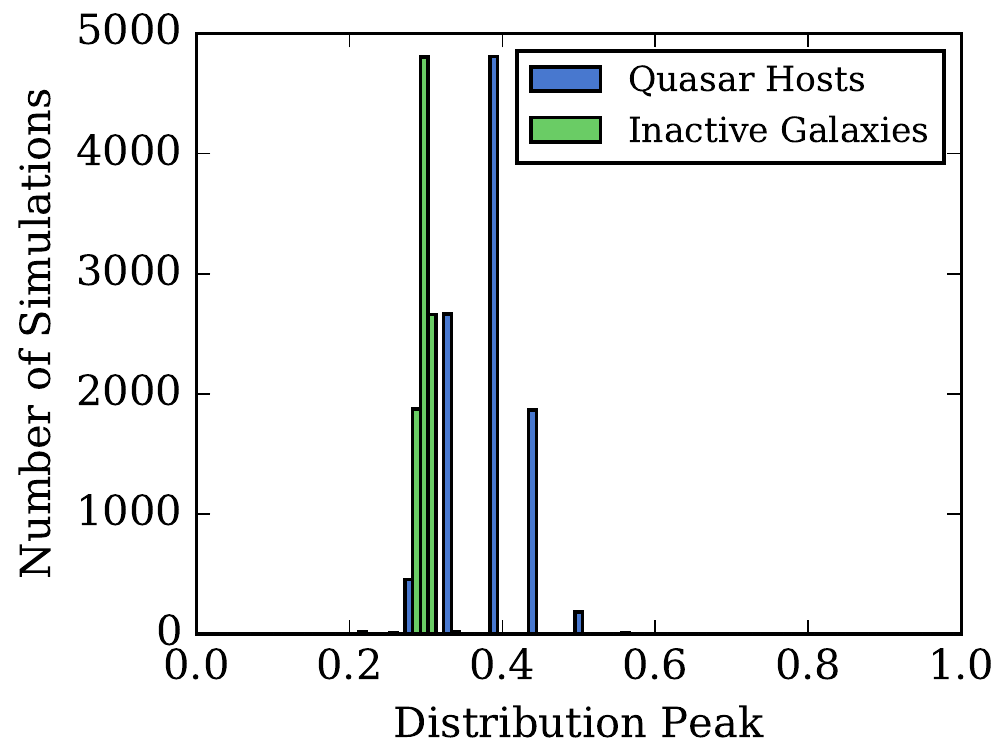}
\caption{Result of 10,000 simulated sequences, to test the precision with which the consensus sequence is determined. Plotted are histograms of the merger fraction distribution peaks. Note that the number of objects is a fixed integer, so the merger fraction distribution peaks can only take discrete rational values. There are many more inactive galaxies, so their merger fraction is primarily determined by the choice of cutoff rank (rank$=$\cutoffrank{}, as in the analysis above). The number of quasars above the merger cut differs by at most one object in 94\% of the simulations.
}
\label{fig:precision_test}
\end{figure}

Another concern is the precision with which the consensus sequence is determined, \ie{} how certain we are of the final ordering of objects in terms of distortion, and whether the uncertainty of this determination could affect the results. This is determined by the precision with which we can estimate the mean rank for each galaxy, approximated by the standard error on the sample mean \change{after excluding non-detections and those removed by sigma clipping (see \sect{expert_ranking}). The standard error on the mean is in turn} determined by the rank variance among experts (sample variance) and the number of classifications for that galaxy. We calculate the standard error on the mean rank for each galaxy, and generate 10,000 simulated sequences, with the merger cutoff performed at rank \cutoffrank{}. \fig{precision_test} shows the result of these simulations. Since there are many more inactive galaxies, their peak merger fraction is primarily determined by the choice of cutoff rank. Roughly 48\% of the simulations result in \emph{exactly} the same quasar merger fraction (the number of quasars above and below the cutoff is unchanged), while for 46\% the number of quasar mergers differs only by one. Thus, while more expert classifiers could reduce the uncertainties on the mean ranks, such an improvement in precision is unlikely to change the qualitative result.

\begin{figure}
\centering
\includegraphics[width=\columnwidth]{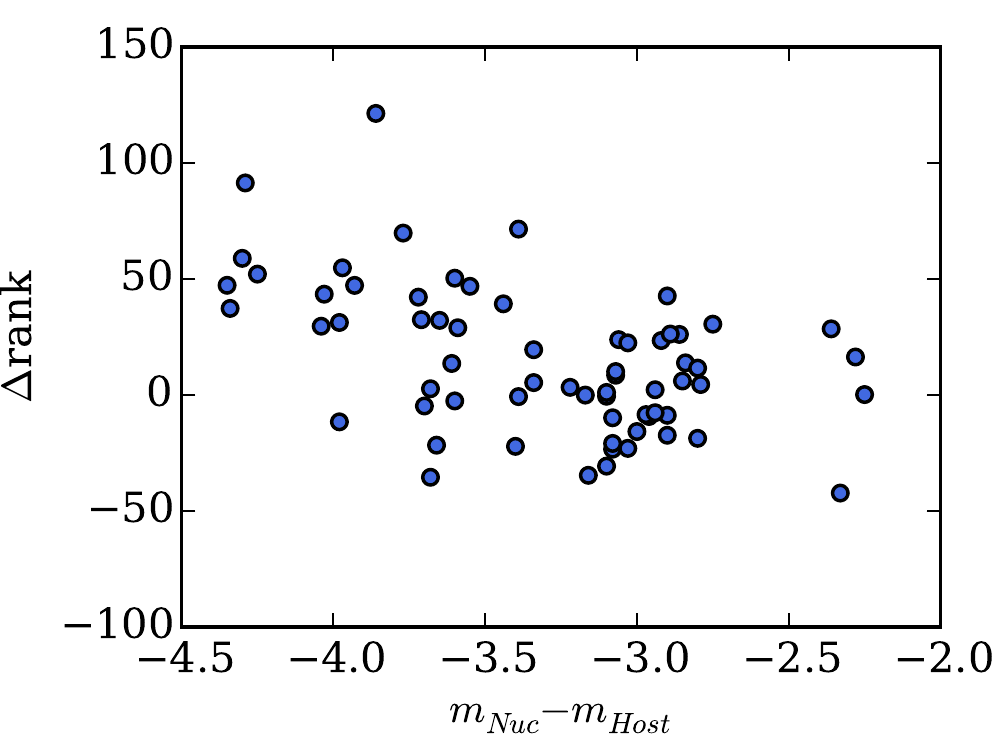}
\caption{Change in rank as a function of nuclear-to-host flux ratio, $m_\mathrm{Nuc}-m_\mathrm{Host}$, for the 36 galaxies that were used as comparisons for more than one quasar, \ie{} had more than one image included in the expert ranking procedure. The x-axis shows the nuclear-to-host contrast ratio, with more extreme contrast ratios on the left. The y-axis shows the change in rank relative to the lowest-contrast image for that galaxy, taken as a fiducial rank, since low contrast ratios have the greatest sensitivity to faint features. The scatter increases with brighter point sources, and there is some bias toward lower ranks (fewer identified merger signatures) at the brightest end. We did not attempt to remove this bias from the consensus ranks, since the inactive sample has the same distribution of $m_\mathrm{Nuc}-m_\mathrm{Host}$ as the quasars by construction, and the bias thus applies equally to both samples.
}
\label{fig:contrast_test}
\end{figure}

Finally, since we had 36 inactive galaxies that were used as comparisons more than once (see \sect{comparison_selection}), we can examine how brighter point sources affect our sensitivity to merger signatures. \fig{contrast_test} shows the change in consensus rank as a function of the nuclear-to-host flux ratio, $m_\mathrm{Nuc}-m_\mathrm{Host}$, relative to the image with the smallest magnitude difference (which has the greatest sensitivity to faint features). There is significant scatter with increasing point source magnitude, as well as some bias toward lower ranks for the brightest point sources (\ie{} experts identified them as having fewer merger signatures). We have not attempted to remove this bias from the consensus sequence, because the same observational limitations apply to both the quasar and the inactive galaxy samples, since by construction the inactive sample has the same distribution of $m_\mathrm{Nuc}-m_\mathrm{Host}$ as the quasars. This underscores our warning against interpreting the merger fractions as absolute major merger fractions, rather than only relative to one another.

\section{Point Source-Subtracted Images and Sensitivity to Faint Features}
\label{sec:appendix_images}
\change{
As discussed in the main text, the presence of a nuclear point source and its subtraction necessarily affect sensitivity to interaction signatures. For \emph{relative} comparisons of the inferred merger fractions between between the two samples --- as in \sect{merger_fraction} and \sect{discussion} above --- it is sufficient that the samples be subjected to the same sensitivity limitations, \ie{} that we added synthetic point sources to the inactive comparison sample, with the same distribution of nuclear-host contrast ratios as the quasar sample. Despite \emph{absolute} merger fraction being a concept with no consistent definition and fraught with problems, some discussion of how the sensitivity constraints differentially affect hosts with distinct morphologies is warranted. \fig{consensus_sequence} shows all 19 quasar hosts and 84 inactive galaxies in consensus sequence order, ranked by co-authors from most to least evidence for major mergers. For inactive galaxies, we show both the images with point source residuals as provided to co-authors for ranking (left panels), and the original CANDELS $F160W$ images for comparison (right panels). 

False positive identifications do not appear to be significant \ie{} point source residuals have not been mis-identified as merger features. Most of the galaxies in the latter half of the sequence (ranks $\gtrsim{}50$), identified as having the little evidence for mergers, are indeed compact and symmetric in the original CANDELS images. However, there are a few galaxies that have stronger evidence for interactions in their original image than is visible in the image used for ranking --- \eg{} ranks 45, 57, 60, 66, and 74. These are disk-like galaxies with a single nucleus and comparatively low surface brightness, but with some large-scale asymmetric features. The asymmetric features are difficult to pick out with the degraded sensitivity after point source subtraction.

Low surface brightness features remain detectable as long as they extend sufficiently far from the removed point source. For example, extended low surface brightness emission is present in SDSS~J124949.65$+$593216.9 (rank 7), SDSS~J131535.42$+$253643.9 (rank 13), and COSMOS~28565 (rank 16), all of which rank highly due to such emission that forms bridges to moderately-bright neighboring galaxies. GOODS-N~30283 (rank 24) appears to be a canonical late-stage major merger with double tidal features --- still visible even in the point source-subtracted image --- though it ranks slightly lower since the double nucleus is not directly visible in the subtracted image. The average surface brightness of the lower tail is $23.8$~\magarc{}, roughly half as bright as the low surface brightness bridges in the luminous, distorted \zsim{2} merger described in \citet{van_dokkum_hubble_2010} and \citet{ferreras_road_2012}.

Thus, while the point source subtraction process does reduce surface brightness sensitivity, it does not wholly conceal the features that are used to identify mergers. For the inactive galaxies, symmetrical galaxies are ranked as having the least evidence for mergers, train-wreck clump chain galaxies are ranked as having the most, and those with slight asymmetries or minor companions fall in the middle, with only a handful of exceptions. Further, any biases against correctly ranking certain morphologies applies equally to both the quasar and inactive galaxy samples. The quasar hosts do not seem to show a preference for any particular morphology compared to the inactive galaxies.

We do not attempt to precisely calibrate an absolute major merger fraction for either sample. However, given the relatively small number of inactive galaxies with wholly obscured merger evidence, we are confident in our assertion that the underlying merger fraction for the quasar hosts is indeed lower than the $\simeq{}0.8$ fraction found in reddened quasar studies (see \sect{comp_red}).
}

\begin{figure\flexstar}
\centering
\newcounter{residfig}
\forloop{residfig}{1}{\value{residfig} < 25}{
\includegraphics[width=0.28\textwidth]{figures/resid_\arabic{residfig}}
}
\caption{Point source-subtracted images of the 19 quasar hosts and 84 inactive galaxies, ordered by consensus rank from most distorted to least distorted. All images use the same asinh stretch, 0\farcs{}060 pixel scale, and $5\farcs{}0\times{}5\farcs{}0$ field of view as the images in \fig{quasar_resids}. \change{Left panels: point source-subtracted images presented to co-authors for ranking. Right panels: original CANDELS $F160W$ images without point source residuals. Quasar hosts have no such image and thus have the words ``Quasar Host'' in its place.} The number in the upper left of each image is its consensus sequence rank (lower numbers are more distorted). Objects which were flagged as non-detections by at least half of individuals were excluded from statistical calculations (see \sect{expert_ranking}), and are annotated with the white text ``Non-detection'' at the bottom of the image. The object names below each image are either the SDSS quasar designation as in \tab{sdss_data} (for quasar hosts), or the field name and galaxy ID number from the 3D-HST catalog (for inactive galaxies).}
\label{fig:consensus_sequence}
\end{figure\flexstar}

\begin{figure\flexstar}
\ContinuedFloat
\centering
\forloop{residfig}{25}{\value{residfig} < 49}{
\includegraphics[width=0.28\textwidth]{figures/resid_\arabic{residfig}}
}
\caption{(continued) Consensus sequence}
\end{figure\flexstar}

\begin{figure\flexstar}
\ContinuedFloat
\centering
\forloop{residfig}{49}{\value{residfig} < 73}{
\includegraphics[width=0.28\textwidth]{figures/resid_\arabic{residfig}}
}
\caption{(continued) Consensus sequence}
\end{figure\flexstar}

\begin{figure\flexstar}
\ContinuedFloat
\centering
\forloop{residfig}{73}{\value{residfig} < 97}{
\includegraphics[width=0.28\textwidth]{figures/resid_\arabic{residfig}}
}
\caption{(continued) Consensus sequence}
\end{figure\flexstar}

\begin{figure\flexstar}
\ContinuedFloat
\centering
\forloop{residfig}{97}{\value{residfig} < 104}{
\includegraphics[width=0.28\textwidth]{figures/resid_\arabic{residfig}}
}
\caption{(continued) Consensus sequence}
\end{figure\flexstar}

\bibliographystyle{apj}
\bibliography{main}

\clearpage
\LongTables

\begin{deluxetable\flexstar}{lcccrrrrr}
\rotate
\tablewidth{0pt}
\tablecolumns{9}
\tablecaption{Quasar Properties in Consensus Rank Order}

\tablehead{
\colhead{Quasar} & 
\colhead{$m_\mathrm{Nuc}$} & 
\colhead{$m_\mathrm{Host}$} & 
\colhead{$M_{V}$} & 
\colhead{$M_\mathrm{BH}$} & 
\colhead{$L/L_\mathrm{Edd}$} & 
\colhead{Rank} & 
\colhead{SEM} &
\colhead{ND Count} \\
\colhead{(SDSS J)} & 
\colhead{(mag)} & 
\colhead{(mag)} & 
\colhead{(mag)} & 
\colhead{$\log{}(\MSun{})$} & 
\colhead{} & 
\colhead{} & 
\colhead{} \\
\colhead{(1)} &
\colhead{(2)} &
\colhead{(3)} &
\colhead{(4)} &
\colhead{(5)} &
\colhead{(6)} &
\colhead{(7)} &
\colhead{(8)} &
\colhead{(9)}
}
\startdata
\begin{filecontents*}{tables/quasars_ranked.csv}
name,mPS,mHost,MHost,MBH,rEdd,rank,sem,ndcount
124949.65+593216.9,18.22,21.48$\pm$0.27,-23.71,9.5,0.35,6,3.8,0
131535.42+253643.9,18.64,20.76$\pm$0.10,-24.18,9.1,0.71,12,7.1,0
155447.85+194502.7,18.94,21.92$\pm$0.16,-23.35,9.7,0.14,17,8.6,0
215954.45$-$002150.1,16.73,20.58$\pm$0.36,-24.44,9.8,0.74,19,6.9,0
232300.06+151002.4,18.19,21.24$\pm$0.21,-23.82,9.7,0.19,21,7.1,0
215006.72+120620.6,18.72,21.13$\pm$0.11,-23.94,9.5,0.29,24,11.1,0
145645.53+110142.6,18.21,20.94$\pm$0.14,-24.19,9.2,0.65,28,7.9,0
120305.42+481313.1,18.24,20.98$\pm$0.16,-24.09,9.4,0.27,33,8.8,0
143645.80+633637.9,17.00,20.41$\pm$0.29,-24.81,10.1,0.22,37,12.7,1
220811.62$-$083235.1,18.53,21.92$\pm$0.23,-23.01,9.2,0.63,38,8.7,0
131501.14+533314.1,18.51,22.26$\pm$0.37,-22.67,10.0,0.10,41,9.2,0
094737.70+110843.3,19.07,21.39$\pm$0.11,-23.51,9.7,0.11,50,14.3,0
081518.99+103711.5,18.97,21.20$\pm$0.13,-23.92,9.3,0.49,66,8.0,0
082510.09+031801.4,17.74,(21.88),(-23.28),9.7,0.28,69,10.6,1
113820.35+565652.8,18.14,21.95$\pm$0.41,-22.97,9.5,0.30,71,11.1,0
085117.41+301838.7,18.88,22.74$\pm$0.34,-22.18,9.0,0.65,72,9.6,0
102719.13+584114.3,18.68,(22.80),(-22.33),9.3,0.57,85,11.6,4
123011.84+401442.9,17.44,19.56$\pm$0.11,-25.63,9.9,0.28,88,11.2,2
135851.73+540805.3,18.25,(22.25),(-22.97),9.4,0.54,95,5.5,9
\end{filecontents*}
\csvreader[no table]{tables/quasars_ranked.csv}{}%
{\csvcoli & \csvcolii & \csvcoliii & $\csvcoliv$ & \csvcolv & \csvcolvi & \csvcolvii & \csvcolviii & \csvcolix \\}
\enddata
\tablecomments{Quasar properties, in consensus rank order. Magnitude values in parentheses indicate $2\,\sigma$ upper limits. Column 1: Quasar name, giving the full sexagesimal coordinates; Column 2: Observed magnitude of the subtracted nuclear point source; Column 3: observed magnitude of the host galaxy; Column 4: Rest-frame $V$-band absolute magnitude of the host galaxy; Column 5: Black hole mass (see \tab{sdss_data}); Column 6: Eddington specific accretion ratio (see \tab{sdss_data}); Column 7: Consensus rank; Column 8: Standard error on the mean of consensus rank; Column 9: Non-detection count, \ie{} number of individuals who flagged the host as a non-detection.
}
\label{tab:quasars_ranked}
\end{deluxetable\flexstar}

\begin{deluxetable\flexstar}{lrrrrrrrrrrr}
\rotate
\tablewidth{0pt}
\tablecolumns{12}
\tablecaption{Inactive Galaxy Properties in Consensus Rank Order}

\tablehead{
\colhead{Field} & 
\colhead{ID} & 
\colhead{R.A.} & 
\colhead{Dec.} & 
\colhead{Redshift} & 
\colhead{$m_{F160W}$} & 
\colhead{$M_{*}$} & 
\colhead{$A_V$} & 
\colhead{Rank} & 
\colhead{SEM} &
\colhead{Use Count} &
\colhead{ND Count} \\
\colhead{} & 
\colhead{} & 
\colhead{(deg)} & 
\colhead{(deg)} & 
\colhead{$z$} & 
\colhead{(mag)} & 
\colhead{$\log{}(\MSun{})$} & 
\colhead{(mag)} & 
\colhead{} & 
\colhead{} & 
\colhead{} & 
\colhead{} \\
\colhead{(1)} &
\colhead{(2)} &
\colhead{(3)} &
\colhead{(4)} &
\colhead{(5)} &
\colhead{(6)} &
\colhead{(7)} &
\colhead{(8)} &
\colhead{(9)} &
\colhead{(10)} &
\colhead{(11)} &
\colhead{(12)}
}
\startdata
\begin{filecontents*}{tables/galaxies_ranked.csv}
field,id,ra,dec,z,mF160W,lmass,Av,rank,sem,usecount,ndcount
GOODS$-$N,16827,189.297348,$+62.225300$,1.97,21.47,11.21,1.3,0,1.0,1,0
AEGIS,38187,214.751160,$+52.830135$,2.13,21.52,11.51,2.5,1,15.0,1,1
GOODS$-$N,3253,189.007736,$+62.150768$,1.98,22.00,10.36,0.4,2,16.5,1,1
COSMOS,28406,150.191299,$+2.480364$,2.02,21.97,10.49,0.7,3,2.3,1,0
UDS,35887,34.379311,$-5.156958$,1.93,21.32,10.95,0.9,4,2.3,1,0
COSMOS,11363,150.119614,$+2.295948$,2.12,21.31,11.07,1.1,5,2.5,3,2
AEGIS,13874,215.097092,$+52.977772$,1.98,21.69,10.27,0.3,7,6.0,1,0
AEGIS,38157,214.643967,$+52.751797$,1.83,21.80,10.97,2.2,8,3.2,1,0
AEGIS,41042,214.597610,$+52.739277$,1.82,21.72,10.76,0.5,9,3.3,1,0
AEGIS,3311,215.045425,$+52.894630$,1.85,21.15,11.33,0.6,10,3.5,2,0
AEGIS,6310,214.780258,$+52.719028$,1.89,21.32,11.12,0.6,11,7.0,2,1
COSMOS,10989,150.077515,$+2.292315$,1.85,21.85,11.00,0.5,13,13.1,1,0
AEGIS,11930,215.078857,$+52.956783$,1.75,21.99,9.90,0.0,14,6.9,1,0
COSMOS,28565,150.139343,$+2.481772$,2.07,21.88,11.40,0.6,15,19.5,1,2
AEGIS,38652,215.093567,$+53.074688$,2.22,22.50,10.43,0.7,16,22.4,1,4
AEGIS,22887,214.666656,$+52.707596$,1.89,22.49,11.11,1.3,18,9.8,1,0
GOODS$-$S,30274,53.131084,$-27.773108$,2.24,21.31,11.16,1.2,20,5.0,3,0
UDS,25091,34.308601,$-5.192824$,1.80,21.34,10.95,0.3,22,5.7,1,0
GOODS$-$N,30283,189.444336,$+62.294060$,2.10,20.61,11.32,1.1,23,5.9,5,17
COSMOS,10592,150.079071,$+2.288243$,1.76,21.73,11.02,0.3,25,9.1,1,0
GOODS$-$N,10657,189.256592,$+62.196178$,2.22,21.58,11.68,1.9,26,5.5,2,0
AEGIS,15871,215.030258,$+52.937485$,2.17,21.91,11.07,1.2,27,5.2,2,0
GOODS$-$N,20709,189.464142,$+62.244041$,1.80,21.64,11.23,0.7,29,5.5,1,0
COSMOS,4536,150.138260,$+2.225018$,1.80,21.89,11.17,2.4,30,6.3,1,0
AEGIS,36755,215.086044,$+53.060520$,1.94,21.63,10.52,0.6,31,10.8,1,0
COSMOS,2716,150.177567,$+2.208040$,1.94,21.79,10.81,1.0,32,10.2,1,0
AEGIS,29591,214.799545,$+52.829071$,1.77,21.96,10.45,1.1,34,8.4,2,0
GOODS$-$S,4568,53.164497,$-27.890385$,2.13,21.37,10.66,0.3,35,9.8,1,0
AEGIS,5745,215.104919,$+52.949425$,1.76,22.48,10.63,0.6,36,16.5,1,2
UDS,922,34.360863,$-5.275831$,1.79,21.93,11.24,0.7,39,7.1,1,0
COSMOS,17406,150.158478,$+2.357966$,1.79,21.34,11.10,1.2,40,8.1,2,0
COSMOS,13083,150.096100,$+2.313470$,2.01,21.76,11.05,0.6,42,5.8,2,1
COSMOS,7884,150.065628,$+2.260996$,2.05,21.27,11.44,0.6,43,8.4,3,0
GOODS$-$S,37243,53.098831,$-27.736591$,1.76,21.42,10.79,1.2,44,12.8,1,0
GOODS$-$S,36095,53.162609,$-27.739000$,1.96,21.18,11.45,0.6,45,9.2,3,0
AEGIS,16065,214.979965,$+52.902481$,2.19,22.51,11.14,2.6,46,9.5,1,0
AEGIS,24059,215.017929,$+52.962994$,1.86,22.54,11.07,1.0,47,10.0,1,0
AEGIS,25526,215.101151,$+53.026985$,1.77,21.94,10.75,0.6,48,12.0,3,12
GOODS$-$S,10562,53.266895,$-27.861710$,1.90,21.76,10.80,0.7,49,8.0,1,0
UDS,14996,34.278858,$-5.226638$,2.07,21.74,11.20,1.3,51,16.3,1,1
AEGIS,29790,215.081818,$+53.030506$,1.92,22.46,9.74,0.2,52,15.5,1,6
AEGIS,24333,215.137421,$+53.046440$,1.97,21.56,11.47,0.9,53,9.9,2,0
AEGIS,25734,214.837219,$+52.840817$,1.88,22.31,10.36,0.9,54,16.6,1,2
GOODS$-$N,2694,189.208069,$+62.146435$,2.10,21.95,10.84,0.4,55,14.1,2,3
UDS,36685,34.289249,$-5.153751$,1.95,21.80,11.29,1.0,56,10.1,1,0
UDS,30475,34.544548,$-5.174858$,1.75,21.24,10.81,0.7,57,9.3,4,9
AEGIS,38153,214.707108,$+52.797298$,1.90,21.83,10.91,2.4,58,14.5,2,0
AEGIS,23768,214.780533,$+52.792133$,1.88,21.84,10.82,1.1,59,9.4,1,0
AEGIS,20962,215.126419,$+53.027435$,2.18,22.51,10.41,1.0,60,9.5,1,0
AEGIS,18678,214.881332,$+52.843845$,1.99,22.34,10.49,1.0,61,13.8,1,2
AEGIS,11730,215.184616,$+53.031445$,1.98,22.48,11.10,0.6,62,10.3,1,1
COSMOS,11136,150.109085,$+2.294313$,1.88,21.89,10.90,2.0,63,10.1,2,0
COSMOS,21413,150.088943,$+2.399108$,1.78,21.77,10.97,1.2,64,7.5,2,0
UDS,23692,34.363182,$-5.199362$,2.16,21.04,11.60,1.1,65,8.0,5,3
COSMOS,7411,150.177017,$+2.255223$,1.79,21.50,11.00,0.6,67,11.3,1,0
UDS,4721,34.239300,$-5.263094$,1.92,21.38,11.40,0.0,68,10.0,1,0
AEGIS,296,215.247437,$+53.020676$,1.98,21.98,10.93,0.9,70,12.3,1,0
UDS,25630,34.372494,$-5.191559$,1.78,20.79,11.26,1.3,73,10.5,4,5
AEGIS,2918,215.168991,$+52.980843$,2.07,21.32,11.53,1.0,74,9.5,3,0
AEGIS,17644,214.906448,$+52.857258$,1.79,21.81,10.99,1.2,75,10.1,1,1
COSMOS,25581,150.154617,$+2.444321$,1.86,21.05,11.31,0.2,76,9.1,5,9
AEGIS,36574,215.094772,$+53.066292$,2.07,21.62,11.19,0.1,77,8.5,2,2
COSMOS,19071,150.087845,$+2.373441$,1.79,21.76,10.82,0.3,78,8.5,1,0
COSMOS,11494,150.073929,$+2.297983$,2.05,20.84,11.57,0.9,79,5.3,4,14
AEGIS,2016,215.226395,$+53.017464$,1.98,22.52,11.00,0.7,80,9.1,1,0
AEGIS,26649,215.042541,$+52.989807$,1.87,21.71,11.13,1.3,81,6.4,2,0
COSMOS,16419,150.095642,$+2.350081$,2.09,21.09,11.29,0.0,82,8.5,5,22
UDS,4701,34.273331,$-5.262073$,2.11,21.11,11.18,0.4,83,8.3,2,3
COSMOS,20983,150.108505,$+2.393802$,1.97,21.66,11.07,0.9,84,9.5,1,0
COSMOS,8443,150.079056,$+2.266573$,1.82,21.11,10.95,0.7,86,6.9,7,19
GOODS$-$N,2295,189.114380,$+62.142494$,2.04,21.78,11.03,0.5,87,6.9,1,8
UDS,32892,34.389614,$-5.168086$,1.90,21.24,10.98,0.3,89,6.9,2,4
AEGIS,9128,214.976212,$+52.872486$,2.13,21.87,11.32,0.5,90,9.5,1,3
GOODS$-$N,33780,189.202026,$+62.317154$,1.86,21.41,11.29,1.1,91,7.8,2,0
UDS,20529,34.549076,$-5.208462$,1.77,21.00,11.26,0.7,92,5.7,5,9
GOODS$-$N,23187,189.514648,$+62.255245$,1.77,21.10,11.08,0.5,93,7.6,6,9
UDS,22802,34.446915,$-5.200699$,1.77,21.05,10.99,0.0,94,8.6,5,9
GOODS$-$N,17735,189.060898,$+62.228977$,1.84,21.48,10.94,0.4,96,8.0,1,1
UDS,19405,34.544777,$-5.211998$,1.86,21.75,10.92,0.4,97,8.3,2,9
AEGIS,12417,215.052643,$+52.940708$,1.82,22.46,10.94,0.3,98,4.5,1,8
AEGIS,35002,215.093201,$+53.058929$,1.79,21.92,11.07,2.2,99,6.8,3,2
AEGIS,12288,215.131424,$+52.995796$,1.98,21.86,10.83,0.4,100,9.9,1,3
GOODS$-$S,42501,53.197029,$-27.712065$,1.82,21.70,11.00,0.6,101,6.4,1,7
COSMOS,17255,150.133347,$+2.355615$,1.76,21.71,10.82,0.3,102,8.8,1,3
\end{filecontents*}
\csvreader[no table]{tables/galaxies_ranked.csv}{}%
{\csvcoli & \csvcolii & \csvcoliii & \csvcoliv & \csvcolv & \csvcolvi & \csvcolvii & \csvcolviii & \csvcolix & \csvcolx& \csvcolxi & \csvcolxii \\}
\enddata
\tablecomments{Inactive galaxy properties, in consensus rank order. Column 1: CANDELS field name; Column 2: 3D-HST catalog ID number; Column 3: Right ascension; Column 4: Declination; Column 5: Redshift (3D-HST \texttt{z\_peak}); Column 6: Observed magnitude in the $F160W$ filter; Column 7: Stellar mass (from 3D-HST SED fit); Column 8: rest-frame $V$-band internal dust attenuation (from 3D-HST SED fit); Column 9: Consensus rank; Column 10: Standard error on the mean of consensus rank; Column 11: Number of quasars for which the galaxy was selected as comparison; Column 12: Nondetection count, \ie{} number of individuals who flagged the host as a non-detection (noting the maximum is Use Count $\times$ 10 experts).}
\label{tab:galaxies_ranked}
\end{deluxetable\flexstar}

\end{document}